\documentclass[useAMS,usenatbib,referee]{mn2e}
\usepackage{graphicx}
\usepackage{natbib}
\usepackage{times}
\usepackage{float}
\usepackage{epsfig}
\newcommand{\bmth}[1]{\mbox{\boldmath${#1}$}}
\newcommand\bb[1]{\bf{#1}}
\newcommand\del{\nabla}
\newcommand\bcdot{\bb{\cdot}}

\newcommand{\cx}[1] {\cal{#1}}

\title[On the formation of a quasi-stationary  twisted disc after a tidal disruption event]{
On the formation of a quasi-stationary twisted disc after a tidal disruption event}
\author[M. Xiang-Gruess, P. B. Ivanov and  J. C. B. Papaloizou ]{ M. Xiang-Gruess$^{1,3}$\thanks{E-mail:mxiang@uni-bonn.de (MXG)}, P.B. Ivanov$^{2}$\thanks{E-mail: pbi20@cam.ac.uk (PBI)} and J. C. B. Papaloizou $^{1}$\thanks{E-mail: J.C.B.Papaloizou@damtp.cam.ac.uk (JCBP)},\\
$^{1}$ DAMTP, Centre for Mathematical Sciences, University of
Cambridge, Wilberforce Road, Cambridge CB3 0WA, United Kingdom \\
$^{2}$Astro Space Centre, P.N. Lebedev Physical Institute, 4/32
Profsoyuznaya Street, Moscow, 117810, Russia  \\
$^{3}$ Helmholtz-Institut f\"ur Strahlen- und Kernphysik, Nussallee 14-16, 53115 Bonn, Germany 
}

\begin{document}

\date{Accepted. Received; in original form}

\pagerange{\pageref{firstpage}--\pageref{lastpage}} \pubyear{2010}

\maketitle

\label{firstpage}

\begin{abstract}

We investigate misaligned accretion discs formed after tidal disruption events that occur when a star encounters a supermassive black hole. We employ the linear theory of warped accretion discs to find the shape of a disc for which the stream arising from the disrupted star provides a source of angular momentum that is misaligned with that of the black hole. For quasi-steady configurations we find that when the warp diffusion or propagation time is large compared to the local mass accretion time and/or the natural disc alignment radius is small, misalignment is favoured. These results have been verified using SPH simulations. We also simulated 1D model discs including gas and radiation pressure. As accretion rates initially exceed the Eddington limit the disc is initially advection dominated. Assuming the $\alpha$ model for the disc, where it can be thermally unstable it subsequently undergoes cyclic transitions between high and low states. During these transitions the aspect ratio varies from $\sim 1$ to $\sim 10^{-3}$ which is reflected in changes in the degree of disc misalignment at the stream impact location. For maximal black hole rotation and sufficiently large values of viscosity parameter $\alpha > \sim 0.01-0.1$ the ratio of the disc inclination to that of the initial stellar orbit is estimated to be $0.1-0.2$ in the advection dominated state, while reaching of order unity in the low state. Misalignment descreases with decrease of $\alpha$, but increases as the black hole rotation parameter decreases. Thus, it is always significant when the latter is small. 

\end{abstract}


\section{Introduction}

A tidal disruption event occurs when
a star approaches  sufficiently close to a supermassive black hole  that it is ripped apart by tidal forces.
Its orbit must take it  within the so-called tidal disruption radius,
$R_T.$ This radius is such that  the mean density of a mass equal to that of the black hole, assumed to be enclosed within a sphere of  radius, $R_T$,
is equal to that of the star. The ensuing tidal disruption results in  an accretion disc around black hole being formed from
the stellar gas.   This in turn gives rise to a luminous source of  radiation.

Over the past two decades or so  around thirty candidate  tidal disruption events (TDE),   where  stars are tidally torn apart  by supermassive black holes  have been identified, see e.g. \citet{Kom2015} for a review and further discussion.
Normally, these are associated with non-stationary flares of soft X-ray radiation in  the centres of previously non-active galaxies, e.g. \citet{Esq2008}.
They are observed to decay  on a time scale  on the order of several years. Some manifest themselves as sources of transient optical line emission thought to  originate from  interstellar gas that has been  ionised by X-ray radiation coming from a TDE \citep[e.g.][]{Kom2008, Kom2009}.
Additionally, there are non-stationary powerful bursts of radiation over a very wide waveband,  ranging from radio to X-rays,  which are interpreted as being produced by processes occurring in a jet directed almost parallel to the line of sight. This is  assumed to be formed during  an  early stage of the  evolution of  the accretion disc formed after a TDE. A canonical example of such an event is SwiftJ1644+57 \citep[see][]{Bur2011}.

TDEs provide an excellent  opportunity to  determine black hole parameters,  study the physics of accretion
and jet formation as well as investigate the nature of stellar populations in galactic centres. Theoretical  investigations of TDEs  were initiated  approximately forty years ago following a seminal paper of \citet{Hil1975}. Subsequently  many researches have considered  aspects of the  formation of TDEs, their properties and observational appearance. These studies can  be  characterised by allocating them to one of  four interrelated groups.

The first of these is  devoted to stellar dynamical processes occurring  in the vicinity of  single or binary
supermassive black holes  which result in  the formation of stellar orbits with their periastrons   close enough  to the black hole for the  stars
 on them to be tidally disrupted.  Tidal disruption rates were evaluated for different parameters characterising the central star clusters and black holes, see e.g. \citet{Fra1976, Lig1977, Mag1999, Sye1999, Iva2005, Sto2016}.
The second group focuses  on the process of tidal disruption itself under different assumptions concerning the  structure of the star, its orbit and the  gravitational field of the black hole \citep[e.g.][]{Car1983, Car1985, Eva1989, Kho1993a, Kho1993b, Iva2001, Iva2003, Iva2006,
Mac2012, Gui2013, Dre2014, Mai2016}. The third group studies the formation, properties and dynamics of the accretion disc formed after a TDE \citep[e.g.][]{Koc1994, Can1990, Kim1999, Sto2012, Hay2013, She2014, Kel2014, Fra2015, Gui2015, Cou2016, Bon2016}. The fourth group explores the  observational consequences of TDEs (e.g. Bogdanovic et al 2004, van Velzen et al 2011, Khabibullin et al 2015, Miller, 2015, Zhang et al
2015).

It was pointed out by \citet{Lac1982}  with further development  by \citet{Ree1988}, that when the stellar orbit is assumed to be parabolic and tidal forces totally  disrupt  the star approximately one half of  the stellar material gains positive orbital energy and is expelled from the system,  while the remainder  attains negative energy (or equivalently positive binding energy) and, accordingly, becomes  gravitationally bound to the black hole.
This  can be seen  if we adopt  a simplified view of tidal disruption as occurring in an abrupt manner  when the star reaches periastron.
At that point one half of the stellar material is situated closer to the  black hole  relative to the centre of mass of the star, while the remainder is further away.
As the centre of mass follows an orbit with  zero binding energy, the stellar material located deeper in the potential well of the  black hole that is moving with the same velocity at the point of disruption will have a binding energy per unit mass equal to the difference between  its potential energy per unit mass and the potential energy per unit mass  at the centre of mass of the star.
The corresponding discussion for material further away than the centre of mass implies that this will become unbound.

Gas elements comprising the bound  material will  in general have  binding  energies ranging between  some largest absolute value and zero.
Assuming their subsequent motion is ballistic, their corresponding orbital periods will be  in the range $P_{min} < P_{orb} < \infty ,$ where $P_{min}$ corresponds to the orbital period of the most strongly  bound material. Accordingly, they return to periastron at different times  after the TDE forming a stream of gas that first arrives at periastron when a time, $\sim P_{min},$ has  elapsed after  the star was tidally disrupted \citep[e.g.][]{Ree1988}.
Supposing  that the amount  of mass occupying  any  small binding energy interval of a fixed extent  is approximately uniform, it is easy to estimate that mass flux  from this stream should be  $\propto t^{-5/3}$ when $t \gg P_{min}$ \citep[e.g.][and referenes therein]{Lod2009}. 
At times order of $P_{min},$ the stream starts to intersect itself near periastron as a result of e.g. Einstein precession, giving rise to the formation of strong shocks.  These shocks convert  stream  kinetic energy into heat, which is later radiated away.

 On the other hand,  at a  sufficiently
early stage of the process,  when friction arising from any effective viscosity will  not have  had enough  time to produce significant effects, its specific  angular momentum remains approximately equal to that of the initial stellar orbit. Thus, there is a tendency to form a gaseous torus  with this specific angular momentum in the vicinity of the  black hole \citep[e.g.][]{Can1990}. On longer time scales  action of an effective viscosity  can cause the torus to  spread,  leading to the formation of an  accretion disc \citep[e.g.][]{Lyn1974}.
Initially, estimates of the accretion rate from the stream indicate that it will be strongly super-Eddington.
The disc  is  expected to be optically thick,  radiation pressure dominated, and, possibly, associated with strong outflows.

When the black hole is non rotating  the mid plane of the accretion disc will coincide with that of the initial stellar  orbit.
However, in the case of  a Kerr black hole, the Lense-Thirring force  acts to drag it to coincide with  the black hole equatorial plane \citep[see e.g.][and references therein]{Fra2015}. On the other hand the stream orbital plane is the same as that of the initial
stellar orbit.  This  is in general expected to be  inclined with respect to the black hole equatorial plane with an inclination angle
order of unity. Accordingly, the stream material arriving in  the region close to the initial  periastron,  after  accretion disc formation
and  assumed relaxation to the equatorial plane,  will impact the disc obliquely, pushing it out of the black hole equatorial plane.
Thus, there is a possibility that  the disc is inclined with respect to the black hole equatorial  plane at radii of  order of the stream impact
radius. This is expected  even in the presence of precession of the stream orbit through some angle produced by  black hole rotation,  the magnitude of this angle is a function of the stellar orbital parameters, the  black hole rotation parameter and  time etc..

In addition, the combined action of black hole rotation and oblique stream impact leads to a non-trivial dependence of  the disc  tilt and twist angle on the distance from black hole.  That the accretion disc is twisted could have a profound effect on its observational properties \citep[see e.g.][]{Bac1999, Cap2007, Wu2010, Dex2013}. This could potentially be used  as a diagnostic for the determination  of the black hole mass and angular momentum as well as to probe the  physical conditions in the accretion flow.

It is the purpose of this paper to  determine the conditions under which  the inclination angle at the stream impact radius can be large as well as to investigate  the  properties of  twisted accretion discs formed after TDE. We tackle the problem using a combination of analytic and numerical techniques. First, we adapt the linear theory of  twisted accretion discs \citep[see e.g.][]{Pap1983, Pap1995}. This formally assumes that the local inclination angle  between  the orbital plane of an annulus of orbiting  disc material  with respect to the black hole equatorial plane is small  throughout the disc
\footnote{ In a situation where the inclination angle at the stream impact radius is of the order of the inclination of the stellar orbit,  as we have indicated,  it is in general expected to be large.  However, we believe that our assumption that it is small does not  affect estimates of the conditions under which this angle will  be significant.}.
In this theory the disc tilt and twist are treated as perturbations on a flat disc
(background) model.  We suppose that the relaxation time  for the disc to attain a quasi-steady twisted tilted
 configuration is smaller than a time scale characterising the evolution of the background flat disc.
As the propagation time for either bending waves or warp diffusion is in general expected to be short compared to the disc evolution time on account of mass redistribution \citep[see][]{Pap1983, Pap1995} for the most part  this  is expected to be reasonable.
Accordingly  we limit consideration to such quasi-stationary configurations. We derive an equation governing  the  tilt and twist of the disc  for which  the stream provides a torque through a source term acting at the  stream impact location.  The magnitude of the torque  is determined by  the magnitude of the inflowing  angular momentum components perpendicular to  the black hole rotation axis  which arise on account of obliquity of the orbital plane of the stream.

We use the governing equation to determine the parameters of the problem that are important  for indicating when there will  be significant misalignment between the disc mid plane
and the black hole equatorial plane. These are found to be a quantity measuring the ratio of the warp diffusion or propagation time to the  local mass accretion time,
together with the ratio of the alignment radius to the stream impact radius.  For radii smaller than the alignment radius, the disc mid plane is significantly modified by the black hole
in the absence of the stream. The dependence on these parameters is investigated using numerical solutions of the governing equation as well as an asymptotic  analytic approach.
In particular we find that large values of the first parameter and small values of the second favour misalignment.

 We test the above approach by performing  three dimensional  numerical  simulations using an  appropriately modified SPH code GADGET-2.
These simulations follow the development of a twisted tilted disc sourced by  a stream produced from tidally disrupted stellar material with the expected range of orbital binding energies.
We find that the approaches obtain the same dependence of misalignment on the black hole rotation parameter with
 a typical difference between disc inclination angles at the stream impact radius,  estimated from the analytic  approach and those obtained from the  numerical simulations,  after an initial relaxation period,  is about  30 per cent, even when the angles is not small, thus validating our general approach.


We go on to study the longer term evolution of the background disc taking account of both gas and radiation pressure using  a one dimensional numerical scheme based on the finite difference code
NIRVANA. This is a practical approach given  the large dynamic range in this problem coupled with the need to consider evolution times greatly exceeding the shorter dynamical time scales present in the system. It is approximate in that a vertical average is performed even though the disc is thick at an early stage when thermal instability leads to  the disc being in a high advection dominated state \citep[see also the slim disc modelling of][]{Abr1988}. We determine the evolution of the  background   accretion  disc model,
incorporating  a mass supply from the stream, through the  advection dominated super-Eddington stage until the beginning of the standard thin disc radiative stage, for different  values of the  Shakura-Sunyaev viscosity parameter $\alpha$.  In the course of this evolution the
disc  semi-thickness, $H$, experiences a very dramatic change from being on the order of the radial scale,  $R$, during the advection dominated stage  down to values order of $10^{-3}R$ at the radiative stage, which has an important consequence for the evolution of the inclination angle.

This transition happens when a typical accretion rate
through the disc at scales of interest is of the order of a few Eddington accretion rates
\footnote{We define the Eddington accretion rate, $\dot
M_E$, as the Eddington luminosity divided by square of speed of light $c$: for pure Hydrogen
 $\dot M_{E}=  {4\pi Gm_pM\over c\sigma_T}$, where $G$ is the gravitational constant, $m_p$ and $M$ are the proton mass and black hole mass, respectively, and $\sigma_T$ is the Thomson crossection.}.
It occurs in an unsteady manner with parts of the disc alternating between high and low states as the accretion rate due to the stream slowly declines.
In this context we  note that in the standard Shakura-Sunyaev model for which  the vertically integrated viscous stress is proportional to the vertically integrated sum of the radiation and gas pressures with the  constant of  proportionality  being $\alpha $, a thermal instability operates when the radiation pressure is larger than that of gas \citep[see e.g.][]{Sha1976}.
 This results in a limit cycle like  behaviour at various locations in the disc after  the transition to the radiative phase begins.
 The transitions are found to occur at  progressively smaller radii as the accretion rate into the disc decreases.
 During such transitions  the disc aspect ratio $\delta = H/R$ experiences a set of transitions between 'low' $\delta \sim 10^{-3}$ and 'high' $\delta
\sim 1$ values until the total pressure in the disc drops down to values such that it  becomes  dominated by that
of gas \citep[see e.g.][]{Szu1997, Szu1998, Szu2001}.

Under our assumption that the disc inclination relaxation time is short compared to
the background disc evolution time  when it undergoes such cycles,  we will see that the cyclic behaviour is reflected in similarly sharp changes of the  disc inclination angle at the stream impact radius between relatively large values at a low state to smaller
 values during a  high state. We use numerical models of the background disc obtained  from the 1D evolution studies
as background models for the quasi-stationary equation describing  the disc  twist and tilt  introduced above, and use this to calculate the evolution
of disc   inclination through the advection dominated stage until the transition to the radiative stage.
\footnote{Note that, strictly speaking  as we assumed the disc to be thin,  our twist equation should be modified when the advection dominated
stage is considered. However, again we assume that this equation can be used during this stage to obtain
approximate  estimates.  In particular the result that the inclination is relatively small then is unlikely to be affected as this is a consequence of the relatively short accretion time
in comparison to the warp diffusion time. A more accurate treatment of the problem will be considered elsewhere.}

We show  that when $\alpha > \sim 0.1$ and black hole rotation is close to  maximal,
typical inclination angles of the disc are of the order
of $\sim 0.1-0.2$ of the stream inclination angle at the advection dominated stage\footnote{At smaller values of
$\alpha $ the angles are  estimated to be even smaller during this stage.} However, the inclination angle becomes larger as the disc aspect ratio decreases,
even in high states, at at later times.
In addition the inclination can grow to values close to  that of the stream during transitions to  low states.
Furthermore, disc inclinations get larger for smaller black hole rotations
at  all stages of the evolution of the disc.


The plan of the paper is as follows. In  Section \ref{Basic} we introduce basic quantities and associated
space and time scales used below. In Sections \ref{gov} -\ref{Numsol} we discuss the equation governing the dic tilt and twist as well as
its solution for a model case with  constant aspect ratio, $\delta $. Section \ref{SPH} is devoted to SPH modelling of the
problem on hand and the comparison between the semi-analytic and SPH approaches. In Section \ref{Grid} we describe
the 1D grid based simulations  and  go on to describe the evolution of  of the background aligned  disc models. We go on to  discuss  solutions to our equation governing
the  disc   inclination
 which incorporates  models obtained from the 1D simulations  as  background models in Section \ref{Gridmodels} . In  Section \ref{edd}  we provide analytic  estimates of
the disc inclination  during the low state for values of the viscosity parameter smaller than those  adopted for  our numerical work. These
are based on an asymptotic analytic theory of solutions of the equation  governing disc  tilt and twist developed in an  appendix.
Finally, in Section \ref{Disc} we discuss our results and set down our  conclusions.

\section{Basic Definitions  and  Notation}\label{Basic}
We  investigate
 the influence of the gas stream produced as a result of a
TDE on the form and structure of an accretion disc around a rotating black hole using both analytic  and
numerical methods.  We envisage the situation where
the entire  accretion disc is formed from the
gas stream resulting from the TDE. At any stage the stream  interacts with a disc
 produced as a result of  the circularisation and subsequent viscous spreading
of stream material that arrived previously.
Thus, we aim  to consider  a fully self-consistent picture assuming that any  accretion disc that was  present
 before TDE had  insignificant mass and so could be neglected.

In the analytic treatment  given bellow we assume  that  the disc  mid plane  is everywhere
close to the black hole equatorial plane, while the  plane  in which the unperturbed stream moves coincides with
the orbital plane  of the disrupted  star and  is  inclined at an angle  $\beta_{*}$
with respect to the black hole's
equatorial plane.   We  introduce a Cartesian coordinate system $(XYZ)$ with origin  at the black hole
location. The  $(XY)$ plane  coincides  with the equatorial plane of the black hole.  The angle between
the $X$ axis and the line of intersection of the plane  containing  the stream  with  the equatorial plane of the black hole is  $\gamma_{*}$.

In the same way we define the inclination  of the disc mid plane  at  radius $R$ to be $\beta(t,R)$ and the
angle between the  line of intersection of this  plane and the  $(XY)$ plane  and the $X$ axis to be
$\gamma (t, R)$.  In the analysis presented below, it is  very convenient to work with the  complex variables
${\cx W}(t,R)=\beta(t,R)e^{{\rm i}\gamma (t, R)}$ and ${\cx W}_{*}=\beta_{*}e^{{\rm i}\gamma_{*}}$. Hereafter we use
caligraphic letters for complex quantities. The angles $\beta$ and $\gamma$ are associated with a tilt
 of the local disc angular momentum vector such that the angle between this vector and the $Z$ axis is $\beta$
 and the angle between its projection on the $(X,Y)$ plane and the $X$ axis is $\pi/2-\gamma.$ This tilt produces a displacement of
a disc fluid element in the $Z$ direction which in the case of small $\beta,$ which will be considered below, 
is equal to $\beta(Y-iX)\exp(-{\rm i}\gamma).$

When the disc is flat the angles $\beta$ and $\gamma$  are obviously constant.  It is  one of the purposes of
this Paper to find the conditions under which  these angles have  a significant dependence on radius and
time due to the influence of stream on the disc.

An important aspect of the problem on hand is that physical processes  arising from  three distinctive phenomena
interact with each other.
These are  associated with the stellar orbit and the stream of gas,  the dynamics of the accretions disc
and relativistic effects determined by the gravitational field of the black hole, respectively.  Accordingly we define
important quantities characterising these three types of processes
and introduce  characteristic
temporal and spatial scales for them  in  turn.
A list of the main parameters and symbols used in this paper is given in Table \ref{table0}.

\subsection{Characteristic spatial and temporal scales and
basic quantities associated with the stellar orbit and gas stream}\label{Char}

We now  specify  spatial and temporal scales associated with the stellar orbit and the gas stream.
For  unit of  distance we use either the periastron
distance $R_p$ of the initial stellar orbit or the distance of the location   where the stream impacts the disc from the black hole, $R_S$.
 In general, we have $R_S > R_p$. When the disc
is sufficiently inclined with respect to the plane of the  stellar orbit as can arise
for a black hole that rotates sufficiently rapidly,  we typically have $R_S \sim R_p$.
These distances  are
expressed  as multiples of the tidal radius, $R_{T},$ with  the multiplication factors  being $1/B_p$ and
$1/B_S$ respectively.
 The quantities  $B_p$ and
$B_S$  are described as penetration factors.
\footnote{In the literature $B_p$ is often denoted by  $\beta$,
see e.g. \citet{Car1983, Car1985}. We use the symbol $B$  to distinguish these quantities
from the inclination angles.}
The tidal radius is given by
\begin{equation}
R_T=\left({{M/ m}}\right)^{1/3} R_*= 7 \cdot 10^{12}M_{6}^{1/3}cm=46M_6^{-2/3}R_g,
\label{eq1}
\end{equation}
where $M$ and $m$ are the masses of black hole and star  respectively,
$R_*$ is the stellar radius,
$M_6=M/10^6M_{\odot}$, and we define the  gravitational radius as $R_{g}=GM/c^2$,
where $c$  and $G$ are the speed of light and the gravitational constant.
In what follows below  we  shall assume that
$m$ and $R_*$ have  Solar values.
We then have  $R_p=R_T/B_p $ and $R_S=R_T/B_S .$

There are several important temporal scales associated with the orbit and the stream. Firstly, there
are the  inverse Keplerian angular frequencies  at the stream impact position and at periastron.
The former is given by
\begin{equation}
t_{S}\equiv \Omega_{S}^{-1}=    {R_S^{3/2}\over \sqrt{GM}}  =   {R_*^{3/2}\over \sqrt{Gm}}B_{S}^{-3/2}\equiv  B_{S}^{-3/2}t_*= 1.6\cdot 10^3B_{S}^{-3/2}s.
\label{eq200}
\end{equation}
and the latter by
\begin{equation}
t_{p}\equiv \Omega_{p}^{-1} =   {R_p^{3/2}\over \sqrt{GM}}  =   {R_*^{3/2}\over \sqrt{Gm}}B_{p}^{-3/2}\equiv  B_{p}^{-3/2}t_*= 1.6\cdot 10^3B_{p}^{-3/2}s.
\label{eq2}
\end{equation}

Secondly, there  is the minimal return time of the stellar material in the stream to periastron after the disruption of the star.  This is estimated as
\begin{equation}
P_{min}={\pi \over \sqrt{2}}(R_p/ R_*)^{3}\left(m/M\right)^{1/2} t_*= 3.5\cdot 10^6M_6^{1/2}B_p^{-3}s.
\label{eq3}
\end{equation}
This is simply the period of an orbit with semi-major axis equal to  $ R_p^2/(2R_*).$
This orbit has a binding energy per unit mass equal to  the change in potential energy per unit mass
experienced when moving  a distance $R_*$ towards the black hole  from pericentre. This gives the  greatest specific
binding energy that material originating from the disrupted star is expected  to have.
After the minimal return time
the disc accretes matter  from the stream at a rate  that can be estimated as
\begin{equation}
\dot M_S={0.5m\over  P_{min}}\left({t\over P_{min}}\right )^{-5/3}={2.8\cdot 10^{26}}B_p^{3}\left({t\over P_{min}}\right)^{-5/3}g/s,
\label{eq4}
\end{equation}
We remark that this follows under the assumption that the mass that has been accreted at any stage is all that was  more strongly
bound than the material currently returning to periastron together with the additional assumption that  there is a linear relation
between the  specific binding energy  of returning  material and the  total mass  of  material that was more  more strongly bound and  has returned previously \citep[see e.g.][]{Ree1988}.

\subsection{Torques acting between stream and disc}\label{Torq}
The unit vector in the direction of the angular momentum  at  a point in the disc expressed in the $(XYZ)$ coordinate system
is given by
\begin{equation}
 {\hat{\bf l }} = (\sin( \beta(t,R)) \sin (\gamma (t, R)),   - \sin(\beta(t,R))\cos (\gamma (t, R)), \cos(\beta(t,R))) \label{unist}
\end{equation}
The corresponding unit vector for the stream is
\begin{equation}
 {\hat{\bf l }}_{*} = (\sin( \beta_{*}) \sin (\gamma_{* }),   - \sin(\beta_{*} )\cos (\gamma _{*}), \cos(\beta_{*} )) \label{unid}
\end{equation}
The rate of  change   of the  component of the disc  angular momentum in the $(X,Y)$ plane  as a result of
interaction with the stream  is
\begin{equation}
\dot {\bf L}=\dot M_SJ_S ({\hat{\bf l}}_{*}- {\hat{\bf k}}( {\hat{\bf l}}_{*}{\bf {\cdot}}{\hat{\bf k}}) )
\label{eq4n}
\end{equation}
where $J_S=\sqrt{2GMR_p} $ is the specific angular momentum of the stream material which corresponds to that associated with
a parabolic orbit with pericentre distance $R_p$ and ${\bf{\hat k}}$ is the unit vector in the $Z$ direction.
We now  assume  that the inclinations $\beta$  and  $\beta_{*}$  are of  small magnitude. Then  with the help of equation  (\ref{eq4n}),
to first order in $\beta_{*}$ we may write
\begin{equation}
\dot {\cx L}=\sqrt{GMR_S}\dot M_S\lambda {\cx W}_{*},
\label{eq4n1}
\end{equation}
where $\dot {\cx L}=i\dot L_X -\dot L_Y .$
Here we have written ${\bf L} = (L_X, L_Y, L_Z)$
 and $\lambda = \sqrt {{2R_p/ R_S}}=\sqrt {{2B_S/ B_p}}$.
To find the  torque acting so as to change the specific angular momentum of the disc material we must
 subtract a contribution corresponding to inserting the stream material with the same specific angular momentum
 as the local disc.  Thus with the help of equation (\ref{unid}),  we  obtain   correct to first order in small quantities
 this torque is given by
\begin{equation}
 {\cx T}= \sqrt{GMR_S}\dot M_S (\lambda {\cx W}_{*} - {\cx W})
\label{eq4n2}
\end{equation}

To  consider further the influence on the disc we
need to specify how the influence of the stream is distributed. Hereafter we assume it is concentrated in
a narrow region around $R_S$ of size $\Delta R_S \ll R_S$, and, accordingly, that consequent torques
can be approximated as being  proportional to
$\delta_{\Delta}~(1~-~R/R_S)/R_S$, where
$\Delta = (\Delta R_S)/R_S.$ Here
$\delta_{\Delta }(x)$ is the so-called 'nascent' delta function.  This is an even function
of $x $ such that  $\int^{+\infty}_{-\infty}dx \delta_{\Delta}(x)=1$, and it converges to the Dirac delta function in the limit $\Delta \rightarrow 0.$

Adopting the above  assumptions  and definitions it is easy to see that we may write
\begin{equation}
 {\cx T}= \int 2\pi\Sigma R\sqrt{GMR} \dot {\cx W}_{*}dR   \label{eq4na}
\end{equation}
where the disc  surface density is  $\Sigma,$ 
the integral is taken over the disc and
 $\dot {\cx W}_{*}$ has the form
\begin{equation}
\dot {\cx W}_{*}={\dot M_S\over 2\pi \Sigma R^2}(\lambda {\cx W}_{*} - {\cx W})\delta_{\Delta}(1-R/R_S).
\label{eq11}
\end{equation}
Note that the parameter $\lambda$ is expected to be unity when the specific angular momentum at the circularisation radius
is the same as that of the stream material. Then we see from (\ref{eq11}) that  $\dot {\cx W}_{*}=0$ when the orbital planes of the
stream and disc are aligned as expected.  We assume, for simplicity, that this condition is valid, and therefore set  $\lambda=1$ in the remainder of the Paper.


\subsection{Basic quantities associated with the dynamics of the  accretion disc}

For our purposes we need to know the evolution of disc aspect ratio, $\delta=H/R,$  with $H$ being the local semi-thickness.
In general,  $\delta$  and $\Sigma$ are both functions of time and radial distance. Immediately after
the disc has been formed it is expected that it will evolve in the advection dominated  slim disc regime and $\delta \sim 1$.
At later stages of evolution, when the accretion rate gets smaller than the Eddington limit  the disc
becomes radiative and $\delta $ is expected to be quite small, $\delta \sim 10^{-3}$.

We adopt the usual
assumption that the evolution of the disc's basic quantities is governed by a turbulent viscosity modelled through
the $\alpha$ prescription, where kinematic viscosity $\nu $ takes the form
\begin{equation}
\nu =\alpha \delta^2 \sqrt{GM R},
\label{eq5}
\end{equation}
where the viscosity parameter $\alpha < 1$ is a constant.
In general, we solve equations determining $\Sigma (t, R)$ and
$\delta (t, R)$ numerically, see Section \ref{Grid} below for a description of our method, and use the obtained
values as inputs for our analytic model for  the evolution of the disc  tilt and twist.
We choose the  unit  of  surface density, $\Sigma_0$, to be determined by the stellar mass and the stream
impact distance  according to $\Sigma_0=m/( 2\pi R_S^2).$
We then use  the  dimensionless surface density $\tilde \Sigma =\Sigma /\Sigma_0$ below.

\subsection{Basic quantities determined by relativistic effects}\label{relativity}

Since the spatial scales we consider are assumed to be significantly larger than the gravitational radius we treat
the influence of relativistic effects through additional effective forces acting in the
classical  Newtonian description.  Accordingly, our SPH simulations are performed using the expression for the
 gravitational acceleration due to  the black hole, ${\bf a}$, given by
\begin{equation}
{\bf a} =-\nabla \Phi + {\bf F}_{GM},
\label{eq6}
\end{equation}
The potential  $\Phi$  is
determined by the Newtonian potential with the addition of a correction which  leads to the apsidal precession of close  free
particle orbits. Thus
\begin{equation}
\Phi=-{GM \over R} - 3 {G^2M^2\over c^2 R^2}.
\label{eq7}
\end{equation}
The form of this correction is chosen so as to provide the same rate of apsidal precession as
 the  expected relativistic Einstein precession \citep[][]{Gar1987}.
 Note that  our hydrodynamical simulations employ  another form of potential proposed by \citet{Pac1980}.
 Although the  Paczynski-Wiita potential gives
the wrong rate of apsidal  precession,  it  gives the correct  radius for  the last stable orbit  in the case of a Schwarzschild
black hole. The reason for these choices was that apsidal precession is potentially significant for a tilted twisted disc at large distances
but less so for an accreting aligned disc for which the location of the last stable orbit may play a more important role.

The gravitomagnetic force per unit mass ${\bf F}_{GM}$ represents the effect of frame dragging \citep[see e.g.][]{Tho1986}.
It takes the form
\begin{equation}
{\bf F}_{GM}={\bf v}\times {\bf B}, \quad {\bf B}=\Omega_{LT}(R)\left({\bf {\hat  k}}-3{({\bf{\hat  k}}\cdot
{\bf R}){\bf R}\over R^2}\right),
\label{eq8}
\end{equation}
where ${\bf \cdot} $ represents the scalar product,  the unit vector ${\bf {\hat k} }$ is along the $Z$ axis of the $(XYZ)$
 coordinate system and ${\bf R}$ is the position  vector such that $R = |{\bf R}|$. The Lense-Thirring
frequency  is given by
\begin{equation}
\Omega_{LT}=2a{G^2M^2\over c^3 R^3}
\label{eq9}
\end{equation}
determines the precession rate of circular orbits of radius $R \gg R_g$  that is slightly inclined to the equatorial plane. It is
proportional to black hole rotation parameter $a$. This  parameter $a$ lies in the range $-1 \ge a \ge 1$, with negative values
corresponding to the situation when black hole rotates in the direction opposite to that of orbital motion. When the black hole is non-rotating $a=0$.

\begin{table}
\begin{tabular}{ll}
 Symbol & Definition   \\
$R, R_{in}, R_{out},  R_*, R_S$ & Radius, disc inner  radius,  disc outer radius, stellar radius, 
and radius to stream impact location\\ 
$  R_p, R_T, R_g,  R_{BP}, R_{rel} $  & Periastron radius,  tidal radius, gravitational radius,
and, alignment scales for  large and small viscosity\\
$\eta, \eta_{rel}$& These are defined by $\eta= R_{BP}/R_S$ and $\eta_{rel} = R_{rel}/R_S$\\
 $B_S, B_p$  &  Penetration factors for stream and periastron\\
 $r, \epsilon,  r_{in}, r_{out}, r_p$& Dimensionless radius,  softening parameter,  and inner and outer boundary radii, and  $r_p=R/R_p.$\\
$\beta, \beta_S, \beta_*$& Inclinations with respect to the black hole  equatorial plane of, the disc, \\
$ $&the disc at the stream impact, and the stream \\
$M, M_6, M_D $& Black hole mass, black hole mass in units of $10^6$  solar masses and disc mass\\       
$ {\dot M}, {\dot M}_S, {\dot M}_E $& Mass accretion rate, mass accretion rate from stream and mass accretion rate at Eddington limit\\
$ m, m_p, {\dot m}, \kappa$ & Stellar mass, proton mass, dimensionless accretion rate and opacity\\
$(X,Y,Z)$& Cartesian  coordinates with $(X,Y)$ plane coinciding with the black hole  equatorial plane\\
$(X',Y',Z'), (R, \theta,\phi)$&Cartesian and spherical polar coordinates with $(X',Y')$ plane containing the stream\\
$\delta, H, \Delta, \delta_{\Delta}$& Aspect ratio, disc semi-thickness, relative radial width of stream input, and nascent $\delta$ function\\
$t, t_*, \Omega_S \equiv t_S^{-1},\Omega_p^{-1} \equiv t_p$ & Time, characteristic dynamical time of star,\\ 
$ $& orbital angular frequencies at the stream impact  location and at periastron\\
${\cal W}, {\cal W}_*, {\cal W}_+ , {\cal W}_-$& Complex inclination $\beta{\rm e}^{{\rm i}\gamma}$ of the disc in general, at the stream location,
and for outer and inner solutions\\
$\Sigma, \Sigma_0$ & General and  characteristic surface densities.  Note that  the dimensionless quantity $\tilde \Sigma =\Sigma /\Sigma_0$\\ 
$\Sigma_1, \Sigma_{high}$ & Scaling parameter for disc surface density and  high state value.  Note that 
$ \tilde \Sigma_{high}= \Sigma_{high}/\Sigma_0$ \\
$k= 3GM/(c^2 \alpha R)$ &Parameter measuring the importance 
of  post Newtonian effects relative to viscosity\\
$\xi = \Sigma \delta^2 R^{1/2} $&  Parameter proportional to viscous mass flux in outer disc\\
$\sigma,\sigma_{max}, \sigma_{rel,max}$& Parameter measuring the ratio of time scales for warp diffusion
 and  local mass accretion,\\ 
$ $&and estimates for this and $\sigma_{rel}=k\sigma$ in the high state.\\
${\bf J}, J_S$&Black hole angular momentum ( $J= |{\bf J}|$) and specific angular momentum of stream\\
  ${\bf L} = (L_X, L_Y, L_Z), T_X, T_Y$ & Disc angular momentum vector and torque components \\
  $\dot {\cx L}=\sqrt{GMR_S}\dot M_S\lambda {\cx W}_{*}$ &  Rate of input of angular momentum due to stream with $\lambda = \sqrt {{2R_p/ R_S}}$ normally equal to unity\\
$\nu, \alpha,\alpha_{crit},\alpha_{-2}$& Kinematic viscosity, viscosity parameter, 
critical value of $\alpha$ below which relativistic effects matter  and $100\alpha$\\
$P_{min},P, P_r, T,  {\cal R}, \mu$& Minimum period, pressure,  radiation pressure, 
temperature, gas constant and mean molecular  weight\\
$\Phi, G, c, {\bf F}_{GM},  \Psi,$& Gravitational potential, Gravitational constant, speed of light, Gravitomagnetic force per unit mass and  phase\\
 $ a, a_R,\sigma_T$& Black hole rotation parameter, St{\'e}fan Boltzmann constant, and Thompson cross section\\ 
$\Omega_{LT},  L_E   $& Lens-Thirring frequency and Eddington luminosity\\
${\hat{\bf l }}, {\hat{\bf l }}_*, {\hat{\bf k }}$& Unit vector in the direction of the disc angular momentum, at the stream
location, and in the $Z$ direction\\
${\bf v}, {\bf R},{\bf v}_{\perp} ={\bmth \omega}\times {\bf R},  {\bf a} , {\bf f}_v$& Velocity, position vector, velocity component perpendicular to ${\bf R}$,  acceleration and viscous force per unit area\\
$v_R, c_s, U, \rho, \langle \rho \rangle  $&Velocity in radial direction, sound speed.  internal energy density, density and  projected density \\ 
$ \epsilon_v,  E,\Pi $ &The rate of energy input per unit area, and   vertically integrated internal energy density and pressure\\  
$t_{sound}, t_{TH},t_{TW}$& Characteristic sound crossing time, thermal time and warp diffusion time at stream impact location\\
$\tau, \tau_{crit}$& Time in units of $P_{min}$ and value of $\tau$ at 
first high state to low state transition (time= $t_{crit}$)\\ 
\end{tabular}
\caption{Table of parameters, variables and symbols.
\label{table0}}
\end{table}

\section{The governing equation for a twisted tilted disc}\label{gov}

In what follows we assume that the disc aspect ratio  $\delta $ and surface density $\Sigma $ change on a timescale that is
much longer than  the timescale associated with evolution of the disc  inclination and  orientation  angles $\beta $ and $\gamma .$
Under this assumption we can solve an  equation for the quantity ${\cx W}=\beta e^{{\rm i} \gamma },$ assuming that a steady state has been set up, in order to find twisted  tilted disturbances of the disc induced by the stream for which there is no explicit dependence on time.

Such  a governing equation has been derived by several authors under various simplifying assumptions \citep[see e.g.][]{Pap1983,
Pap1995, Iva1997, Dem1997}. In this paper we adopt the form  obtained by \citet{Iva1997}, hereafter II, who assumed that $\beta, $  $\alpha, $
$\delta $ and the ratio of gravitational radius to a radius  of interest are all small. This takes the form
\begin{equation}
{\delta^2 \sqrt{GM}\over 4\alpha \xi R}{d\over dR}\left(\xi R^{3/2}{(1+ik)\over (1+k^2)}{d{\cx W}\over dR}\right)+i\Omega_{LT}{\cx W}
+\dot {\cx W}_{*}=0,
\label{eq10}
\end{equation}
where
\begin{equation}
k={3GM\over c^2 \alpha R}
\label{eqn10}
\end{equation}
determines the contribution of post-Newtonian corrections to the equation of motion.
The last term in (\ref{eq10}) describes the influence of the stream on the disc.
It is given by equation (\ref{eq11}) and is absent in II. { The factor $\xi= \Sigma \delta^2 R^{1/2}$ is approximately
constant for a 'standard' thin accretion discs sufficiently far from the last stable orbit. For such a disc
the  inward advective flow of angular momentum is approximately balanced by outward angular momentum flow transferred
by viscous forces, and, in this case $\xi$ is proportional to the flux of mass in the disc. In our case there is
a mass inflow in the disc due to the presence of the stream, and the disc is, in general, time-dependent and may
be moderately thick. Thus, $\xi $ is retained in our numerical solutions of (\ref{eqn10}), where both $\delta $ and
$\xi$ are taken from a background numerical model, see below. However, we set $\xi = const $ in our analytic estimates
for simplicity.

Note that (\ref{eq10}) implies time dependence only implicitly, through the factors $\delta$, $\xi$ and $\dot {\cx W}_{*}$,
which are, in general, functions of time. This is approximately valid when characteristic time scales associated with
the evolution of disc's tilt and twist are much smaller than the ones corresponding to the background quantities and
$\dot {\cx W}_{*}$. This condition may be broken in course of evolution of our system, especially during the transition
of disc from 'high' state with a large $\delta \sim 1$ to a low state with $\delta \ll 1$, when the thermal instability
may operate on a relatively short time scale $t_{TH}\sim \alpha^{-1}t_{S}$ and sharp features in distribution of
$\xi $ and $\delta $ may propagate over the disc on the sound crossing time scale $t_{sound}\sim \delta^{-1}t_{S}$.
When $\alpha > \delta$  both $t_{TH}$ and $t_{sound}$ are smaller than a characteristic time scale of tilt and twist diffusion
$t_{TW}\sim ( \alpha/ \delta)t_{S}.$  In the opposite limit tilt and twist have a typical propagation time
order of $t_{sound}$ and dissipation time scale on order of $t_{TH}$. Clearly, the assumption of a stationary twisted disc
may not be  valid when it  is undergoing rapid transitions in either case. However, the stationary states  represent the target states  that the disc evolves towards
at any time and so we shall make  the  assumption in this Paper in order to obtain  an estimate of
typical possible disc inclinations and their dependence on the  parameters of the problem as a first step towards  constructing more
realistic time dependent models of a twisted disc under the influence of  a gas stream.}

All the terms in (\ref{eq10}) correspond to  projections of  torques induced in the disc onto the equatorial
plane , divided by the value of Keplerian angular momentum stored in the disc per unit radius which is given by
${dL/ dR}=2\pi \Sigma R \sqrt{GMR}$. These projections enter (\ref{eq10}) in combinations ${\cx T}~=~iT_X~-~T_Y$,
where $T_{X,Y}$ are  $X$ and $Y$ components of the torques (see for example Section~\ref{Torq}).

It is convenient to introduce dimensionless radial distance $r=R/R_S$ and rewrite (\ref{eq10}) in the
form
\begin{equation}
{1 \over r\xi }{d\over dr}\left(r^{3/2}\xi {(1+ik)\over (1+k^2)}{d{\cx W}\over dr}\right) \pm i{\eta^{3/2}\over r^3}{\cx W}
+\sigma ({\cx W}_*-{\cx W})\delta_{\Delta}(r-1)=0,
\label{eq12}
\end{equation}
where the positive  (negative ) sign corresponds to  prograde (retrograde) rotation of the  black hole with respect to
the disc gas, we have used  equation(\ref{eq11}) and ${\cx W}_*=\lambda {\cx W}_{*}.$
\begin{equation}
{\rm In\hspace{1mm} addition\hspace{2mm}}\eta=R_{BP}/R_{S}={2\over 23}B_S(\alpha |a|M_6\delta^{-2})^{2/3} \quad {\rm with} \hspace{2mm} R_{BP}=4(\alpha |a|)^{2/3}\delta^{-4/3} {GM\over c^2},
\label{eq13}
\end{equation}
and
\begin{equation}
\sigma={4 \alpha \over \delta^2}{\dot M_S\over 2\pi \Sigma R_S^2}\sqrt{{R_S^3\over GM}}.
\label{eq14}
\end{equation}

\subsection{Important parameters determining the response of the disc to the incoming stream}\label{important}
\noindent  The quantity $\sigma$ can be simply  interpreted as the ratio of the time  scale for warp diffusion over the length
scale $R_S$ to the local mass accretion time scale. Its value is accordingly   expected to be  important
for  determining the expected disc misalignment produced by a misaligned stream.
A large value implies warp propagation should be  ineffective at dispersing  accreting misaligned angular momentum
and thus favour a misaligned disc.
\noindent Note in addition that $R_{BP}$ characterises the scale of disc alignment with the equatorial plane of the black hole at small radii in the absence of the stream. This
always occurs  when the relativistic correction parameter  $k$ may be neglected. The parameter $\eta =R_{BP}/R_S$  measures the importance of the tendency towards alignment
at the stream impact radius and it is therefore its value is important for determining the disc response, a large value favouring alignment.
The tendency towards alignment with the black hole equatorial plane   is known as the Bardeen-Petterson effect \citep[][]{Bar1975}.

However,  this  does not operate in the disc when
$\alpha < \alpha_{crit}=|a|^{-2/5}\delta^{4/5}$
 and the black hole rotation is
prograde, $a > 0$, see II97. In that  case, the relativistic correction parameter, $k,$  is important and there are oscillations of the inclination angle at small radii
instead of  alignment \citep[see also][]{Dem1997, Lub2002, Zhu2011, Mor2014, Zhu2014}.  In the  case of retrograde rotation  alignment  can take place at all reasonable  values of $\alpha$.
When $\alpha < \alpha_{crit}$   the alignment scale
differs from $R_{BP}$,  and it does not
depend on $\alpha$ (see II).
This typical radial scale, $R_{rel}$, together with the ratio $\eta_{rel}=R_{rel}/R_S,$ which plays the role of $\eta$ in this case,  can defined as
\begin{equation}
R_{rel}=4|a|^{2/5}\delta^{-4/5}{GM\over c^2}, \quad {\rm with} \quad \eta_{rel}=R_{rel}/R_S={2\over 23}B_S |a|^{2/5}M_6^{2/3}\delta^{-4/5}
\label{eqn16}
\end{equation}

\subsection{Numerical solutions of the governing  equation } \label{Numsol}

\subsubsection{Boundary conditions and solution method}
In general, equation (\ref{eq12}) should be solved numerically. In order
to do that we need to  specify the inner and outer boundary conditions.
As the outer boundary condition at some outer radius of the computation
domain $r_{out}$ we adopt  ${d }{\cx W}/dr =0$ to mimic a regularity condition at the disc outer edge.
{In general, the inner boundary condition
set at an inner radius of the computation domain, $r_{in}$,  should be
be different  according to whether  solutions of the homogeneous form of (\ref{eq12}) ( for which  ${\cx W}_*=0$ )  possesses growing and decaying modes (i.e. when either $a > 0$ and $\alpha  > \alpha_{crit}$ or  $a < 0$) or when the solutions are  oscillatory
( $a > 0$ and $\alpha < \alpha_{crit}$).

However, in the former case a precise form of the inner condition is actually not important.
Indeed, when the equation is numerically integrated from $r_{in}$ towards larger radii its solution rapidly converges to the growing mode regardless of the form of the inner boundary condition.
In the latter case, setting a different
boundary condition would only lead to a different phase of the oscillations, which shouldn't
influence our qualitative results.  Therefore, in this
Paper, for simplicity,  we  adopt  ${d }{\cx W}/dr =0$ at the inner radius of computational domain
as well. Note that such inner and outer boundary conditions ensure that disc's inclination vanishes
when the forcing term due to the stream disappears. }

In order to obtain  a solution with the specified inner and outer boundary
conditions we employ a fitting point procedure.  To do this we solve both the homogeneous and inhomogeneous forms
of (\ref{eq12})  starting from $r_{in}\ll 1$ and ending at $r=1$, and also
starting from $r_{out} \gg 1$ and ending again at $r=1$.  We  then choose multiplication  coefficients for  the homogeneous
solutions  which are such as  when these are added in,  the requirement of continuity of the solution and its radial derivative at $r=1$ is satisfied.
When such procedure is used it is evident that a precise form of inner and outer boundary conditions does not play a significant
role in all cases for which  growing and decaying modes exist, since they are singled out by the numerical procedure.

 \subsubsection{Numerical results for constant  $\delta$ }

Solutions of (\ref{eq12}) using time dependent background numerical models based on either SPH or finite difference
scheme are discussed in the subsequent Sections. Here we  consider models with constant
$\delta $ and $\xi$ as they   provide  qualitative information on the possible behaviour of more complicated cases.

\begin{figure}
\begin{center}
\vspace{10cm}\includegraphics{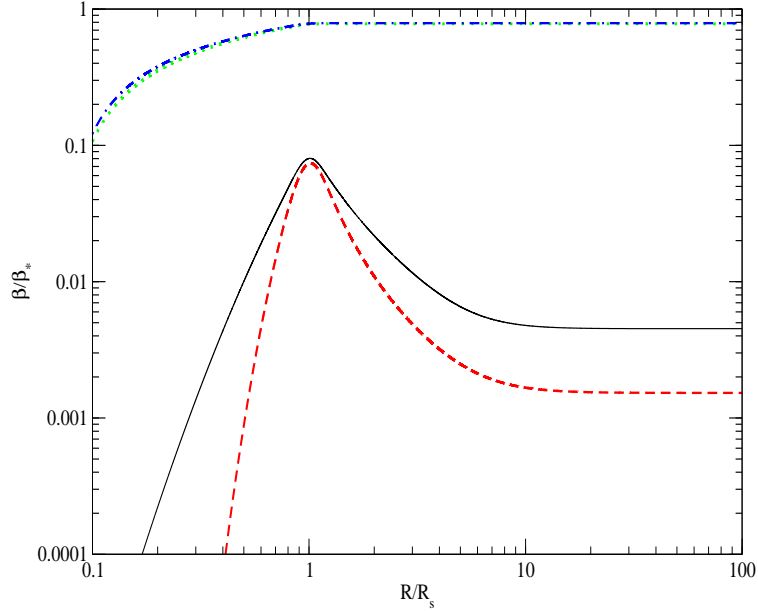}
\end{center}
\vspace{-0cm}
\caption{The dependence of inclination angle $\beta $ in units of
its maximal  $\beta_*$ on radius is shown.
  Cases  with  $\alpha > \alpha_{crit}$ and $\sigma=1$  are illustrated.
The  parameters corresponding to the  different curves are given in the text.}
\label{Fig1}
\vspace{-0.0cm}
\end{figure}

\begin{figure}
\begin{center}
\vspace{12cm}\includegraphics{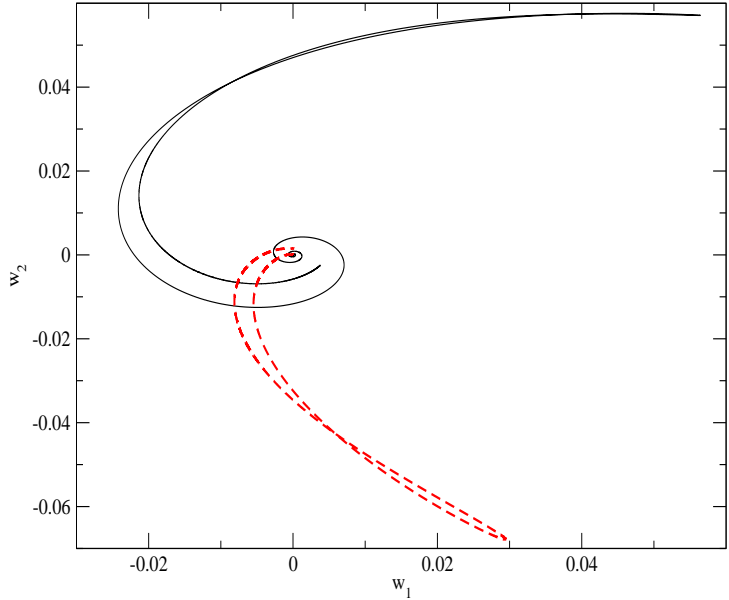}
\end{center}
\vspace{0cm}
\caption{A parametric representation of the solutions corresponding to $\eta=8.7$
shown as curves
on the plane $(w_1=Re({\cx W}), w_2= Im({\cx W})$. Solid and dashed curves
show the cases with  $a > 0$ and $a < 0$, respectively. For more details see the text.}
\label{Fig2}
\vspace{-0.0cm}
\end{figure}

\begin{figure}
\begin{center}
\vspace{12cm}\includegraphics{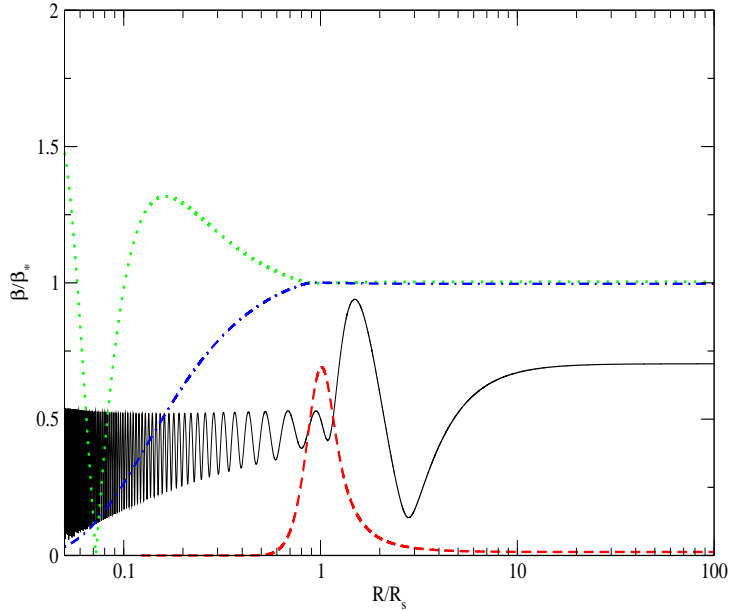}
\end{center}
\vspace{0cm}
\caption{As in  Fig. \ref{Fig1} but for cases with $\alpha < \alpha_{crit}.$
The  parameters corresponding to the  different curves are given in the text.}
\label{Fig3}
\vspace{-0.0cm}
\end{figure}

Solutions of (\ref{eq12}) obtained under the  assumptions stated above
are shown in Figs \ref{Fig1}-\ref{Fig2} for the case with
$\alpha > \alpha_{crit}$ and in Fig. \ref{Fig3} for the case with $\alpha < \alpha_{crit}$.
In the former case we present  solutions  representing each of the four  regimes
which correspond to positive and negative values of $a$ together with
values of $\eta $ greater and less than unity.
In the latter case an analogous set of  four cases, defined as in the former case  are considered,
the only difference being that $\eta $ is replaced  by $\eta_{rel}.$
In all of these  cases we set $\sigma=1.$  We go on to
discuss the dependence of a typical inclination angle on values of $\sigma $ below .


In Fig. \ref{Fig1} the inclination angle $\beta $ is shown in units of
its maximal theoretically expected  value $\beta_*=|{\cx W}_*|$  as
a function of $r$, for  cases when $\alpha > \alpha_{crit}$.
 As  $\lambda=1,$  $\beta_*=|{\cx W}_*|$ corresponds to the case
where the disc and stream orbital planes are aligned.
 In the absence
of warp or twist propagation, the stream is not expected  misalign
the disc with respect to the equatorial  plane
of the black hole  to a greater extent.

Solid and dashed lines respectively correspond to prograde and retrograde black hole rotation.
 For these cases  $\alpha=0.1$, $\delta=0.01$,  $|a|=1$, $M_6=1$, accordingly, $\eta=8.7 > 1$ and $\eta_{rel}=3.5$.
Dotted and dot dashed lines show solutions for prograde and retrograde rotation
for the case of small $\eta=0.2$.
The  other parameters for these are $\alpha=0.5$, $\delta=0.056$, $|a|=1$, $M_6=0.02$
 and $\eta_{rel}=6.4\cdot 10^{-2}$.
One can see that when $\eta $ is fixed the
curves corresponding to $a > 0$ and $a < 0$ are almost indistinguishable.
 For both large and small $\eta$ the disc aligns
  with the equatorial plane at small $r$.
  However, when $r > 1$ these cases behave differently.
  When $\eta $ is large the inclination angle drops to values much smaller than the
  maximal value at $r=1$ at larger radii. On the other hand for small $\eta,$
   the inclination angle is approximately constant at $r > 1$.
   These results simply represent the effect of the alignment radius moving outwards as $\eta$ is increased
   and so causing the disc to align at larger radii.

Fig. \ref{Fig2} shows the trajectory of the prograde and retrograde
solutions having $\eta=8.7$
on the plane  $(w_1=Re({\cx W}), w_2=Im({\cx W})$ with $r$ as  parameter.
Note that for small inclinations the radial and angular polar coordinates
corresponding to $w_1(r)$ and $w_2(r)$ are  $|\beta|$  and $\gamma$ respectively.
Both curves have their origin at $r = r_{min}$. When $r $ grows, but is less than  unity,
the curve corresponding to $a > 0$ ($a < 0$) spirals clockwise (counterclockwise), but
the direction of evolution changes when $r$ becomes greater than unity. That means
that when $r < 1$ the rotation of the polar angle $\gamma$  is always directed in the
sense of the black hole rotation, see also II and ZI.

Fig \ref{Fig3} shows the evolution of the inclination angle
   when $\alpha < \alpha_{crit}$  and the relativistic
correction $k$ determines the shape of the disc. Solid and dashed curves
represent the case of large $\eta_{rel}=5 > \eta=1.72$ for prograde and retrograde
rotation of the black hole respectively.
Other parameters corresponding for these cases
are $\alpha=3.5\cdot 10^{-3}$, $\delta=6.3\cdot 10^{-3}$,
 $|a|=1$ and $M_6=1$.
 Dotted and dot dashed curves illustrate cases with
 relatively small $\eta_{rel} = 0.2 > \eta = 3\cdot 10^{-3}$
for  prograde and retrograde black hole rotation respectively. Other parameters were
$\alpha=3.5\cdot 10^{-3}$, $\delta=6.3\cdot 10^{-3}$,  $|a|=1$ and $M_6=1$.
The most important difference between these cases and those with $\alpha
> \alpha_{crit}$ is that the disc does not align at small radii when the
black hole rotation is prograde. Instead there are radial
oscillations of the inclination angle with  wavenumber  and amplitude
increasing towards black hole, see II.. This effect is much
more prominent in the cases with large $\eta_{rel}$. Note, however, that in this
case the approximations leading to (\ref{eq12}) fail to be valid at radii
$r \ll \eta_{rel}$ and, therefore, oscillations  in the disc inclination with radial
wavenumber  $\gg 1/R_{rel}$ are likely to be absent.
We remark that  values of $\beta $ at $r=1$
are approximately the same for curves corresponding to the same $\eta_{rel}$.

We now go on to make a comparison of numerical solutions to (\ref{eq12})
with  those obtained using a simple analytic approach developed in Appendix A.
The cases having $\eta=8.7, 0.2$ and
$\eta_{rel}=5,0.2$ will be  hereafter refered to as cases 1-4, respectively.
The appropriate values of $\alpha$, $\delta $ and
$M_6$ are specified above.

\begin{figure}
\begin{center}
\vspace{12cm}\includegraphics{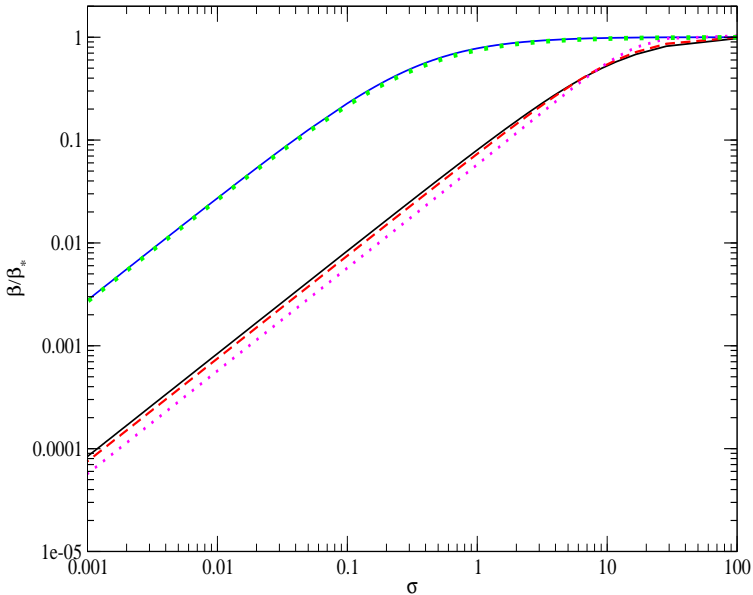}
\end{center}
\vspace{-0cm}
\caption{The dependence of  the disc inclination  at the stream impact position, $\beta_S$, is shown as a function of the parameter $\sigma$ for
two disc's models specified above as cases 1 and 2, where $\alpha > \alpha_{crit}$.  The solid line with a smaller value of $\beta_S$ at a
given $r$ and the dashed line represent numerical solutions of
(\ref{eq12})  corresponding to case 1 with  $a=1$ and $a=-1$, respectively.  The dotted curve  with a smaller value of its argument is for the
same case, but calculated with help of equation (\ref{eqn22}). Solid and
dotted lines with larger values of their arguments  are for  case 2.  The solid line is calculated numerically, while the dotted one is given by equation
(\ref{eqn23}). Both lines correspond to $a >0$.}
\label{Fig4}
\vspace{-0.0cm}
\end{figure}

In Fig. \ref{Fig4} we show a comparison of our analytic approach with numerical
solutions of equation (\ref{eq12}) having $\alpha > \alpha_{crit}$.
We plot the dependence of  the disc inclination  at the stream impact position, $\beta_S$
in units of $\beta_*$ as a function of the parameter $\sigma.$
For the case
with  $\eta=8.7 > 1$ (case 1)  we show two numerical curves corresponding to  prograde and retrograde black hole rotations. A typical deviation of these curves
from the result following from equation (\ref{eqn22}) is about 30 per cent.
For the  case with small $\eta=0.2 < 1$ (case 2)  curves corresponding to prograde
and retrograde rotations practically coincide, therefore only the one
with  $a > 0$ is shown. The analytic result following from equation (\ref{eqn23}) is also very close to the numerical curves.
We remark that  $\beta_S$ becomes significant for large $\sigma.$
As $\sigma$   is a measure of the warp diffusion time to the accretion time, this is as expected.

\begin{figure}
\begin{center}
\vspace{12cm}\includegraphics{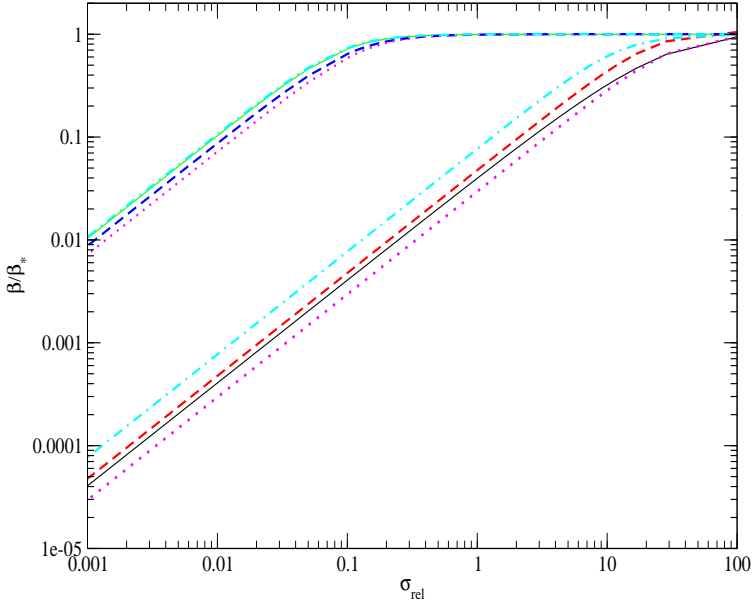}
\end{center}
\vspace{-0cm}
\caption{Same as Fig. \ref{Fig4}, but for the cases 2 and 3 having $\alpha < \alpha_{crit}$  and $\beta_S$ is shown a function of $\sigma_{rel}=k\sigma$, see the text for a description of particular curves.
}
\label{Fig5}
\vspace{-0.0cm}
\end{figure}

In Fig. \ref{Fig5} we show results of a  corresponding comparison between numerical results for the cases 3 and 4 with $\alpha < \alpha_{crit}$ and those obtained
from our  analytic expression (\ref{eqn20}). Here
we plot the dependence of  the disc inclination  at the stream impact position, $\beta_S$
as a function of the parameter $\sigma_{rel}=k\sigma.$

The uppermost curves of a given type correspond to
 case 3 for which $\eta_{rel}=5$ is relatively large, while lowermost curves of a given type correspond to  case 4 for which
$\eta_{rel}=0.2$ is small. Solid and dashed lines respectively  show numerical results for prograde and retrograde rotation, while dotted and dot dashed lines  are obtained analytically for $a=1$ and $a=-1$. One can see  that the disagreement between numerical and analytic results in this case is now  of  the order of 30-40 per cent.  This means that the simplified  analytic solutions  are accurate enough   for our purposes.

{Note that so far we have considered only solutions corresponding to the case of constant $\delta $. In our numerically obtained models of
background quantities two qualitatively different dependencies of $\delta $ on $r$ arise during a period of transition from a 'high'
to 'low' state. Since this transition initially occurs at larger radii, there are configurations with $\delta $ being order of unity
at radii smaller than some 'transition' radius $r_{tr} > 1$, and $\delta \sim 10^{-3}$ for  $r > r_{tr}$. When the transition
happens at radii  of $r \sim 1$  there appears distributions of $\delta $ in the low state $\delta \sim 10^{-3}$ at radii both smaller
and larger than $1$, while at $r \sim 1,$ $\delta $ can be significantly larger than its typical 'low' value due to heating of the disc gas
through the input  of kinetic energy carried by the stream. Such configurations have an intermittent nature through the  development of thermal
instability,  provided that radiation pressure dominates over the  disc  gas  pressure and the $\alpha $ prescription for viscosity is assumed
to be valid. In principle, they can also be treated by a technique similar to  that  used in Appendix A for the case of constant $\delta $. However, here, for simplicity, we do not consider these more complicated cases leaving them for a future work. }

\section{SPH Simulations of discs supplied by a stream resulting from tidal disruption}\label{SPH}

We  have performed SPH  simulations  of the evolution of a disc which is impacted by a gaseous stream resulting from tidal
 disruption of a star. We use  a modified version of the publicly available code {\rm GADGET}-2  \citep[][]{Spr2005}.
  This is a hybrid N-body/SPH code capable of modelling both fluid and distinct massive bodies that interact with it.

In our case we incorporate a rotating  black hole of fixed  mass $M.$
Relativistic effects  are incorporated by  adding  effective forces acting within a
classical  Newtonian description (see Section \ref{relativity}).
We adopt spherical polar coordinates $(R, \theta,\phi)$ with origin at the   location of the black hole.
 The associated Cartesian coordinates   $(X',Y',Z')$  are chosen such that the $(X',Y')$ plane coincides with initial
 orbital plane of the tidally disrupted star. In addition
the location of the $Y'$ axis is chosen such that  the angular momentum  vector of the black hole is  given by
  ${\bf J}=(0, J \sin(i), J \cos(i))$, where $J = a GM^2/c,$
 with $a$ being the black hole rotation parameter and   $i$ is  the inclination of this vector to the $Z'$ axis.
 We recall that ${\bf J}$  defines the direction of the $Z$ axis for  the coordinate system with $(X,Y)$ plane
  coinciding with the equatorial plane of the black hole that we adopted in Section \ref{Basic}. In this Section for convenience
 we shall choose the $X$ axis to coincide with the $X'$ axis which
 corresponds to $\gamma_*$ defined in Section \ref{Basic} taken to be zero. 
For our simulations, we take the penetration factor $B_p=5/3,$  $i =\pi/4$ and $M=10^6 M_\odot$.

The   dimensionless time unit for these calculations was taken to be
  the  inverse periastron frequency $\Omega_p^{-1}$ and in this Section
the   dimensionless unit of length  is taken to be  the periastron distance $R_p,$ thus $R$ expressed in 
 dimensionless units  is $r_p = R/R_p.$.

\noindent  The gaseous disc  and  the stream are  represented by SPH particles.
The total unsoftened gravitational potential $\Phi$ at a position ${\bf  R}$ is given by  equation (\ref{eq7})
and the gravitomagnetic force per unit mass by equation (\ref{eq8}).
An important issue  for N-body/SPH simulations is the choice of the gravitational softening lengths.
 The only gravitational softening that is included in our simulation applies to
 the gravitational interaction between the SPH particles and the black hole.
For the   practical computation  of the gravitational interaction between the black hole and the gas particles,
  the potential and gravitomagnetic force  were softened following the method of \citep{Spr2005}.
This was implemented with  fixed  softening length $\varepsilon=0.05$ in  dimensionless units.
Shocks were  handled following the procedure of \citet{Spr2005}. 
In particular the parameter  $\alpha$ that scales
the magnitude of the applied viscous force  that is defined in equation (14)
of \citet{Spr2005}   (but  which is not used in that context elsewhere in this paper)  
 was chosen to be 0.5.
 Furthermore the black hole  was assumed to  accrete  gas particles that approach it to  within  0.1 dimensionless  units.
Thus such particles were removed from the simulation.
In addition the gravitational effect of the disc on the black hole is neglected.


\subsection{Initial conditions} \label{sec:IC_SPH}
\subsubsection{Disc setup }
The disc  setup is such that the angular momentum vector for all particles was in the same direction enabling a midplane for the disc to be defined. This midplane is set up such that the disc's total angular momentum vector  is parallel to the magnetic spin vector of the black hole.
The particle distribution  was chosen to  model  a  disc with surface density profile  given by
\begin{eqnarray}
\Sigma=\Sigma_1 R^{-1/2}. \label{eq:sigma}
\end{eqnarray}
Here $\Sigma_1$ is a constant
 \noindent The disc mass is then given by
\begin{eqnarray}
M_D=2\pi \int_{R_{in}}^{R_{out}} \Sigma(r')r' dr'=\frac{4}{3} \pi \Sigma_1 R_{out}^{3/2}\ ,
\label{discmass} \end{eqnarray}
where $R_{in}$ and $R_{out}$ are the inner and outer disc boundary radii.
When these and the disc mass  are specified (\ref{discmass}) is used to determine $\Sigma_1.$
For the simulations presented here, we adopted $M_D = 0.2m$,  $R_{in}=0.5$ and $R_{out}=1$ 
with the last two being given  in dimensionless units.
 As self-gravity is expected to play a minor role, it is neglected in the simulations. The disc was evolved for several hundred time units
 in order to attain a relaxed quasi-steady distribution before being allowed to interact with a mass stream.

\subsubsection{Setting up the star at pericentre and generation of the mass stream}
We adopt a simple procedure for generating a stream generated  that might be expected to arise from a tidally disrupted star that
subsequently provides a mass source for a disc that is formed partly from preexisting
material as well as that from the stream.
The particles comprising  a 'star' are initially  set up  so as to form  a homogeneous  sphere of radius $R_\odot$
with its centre of gravity at  the  pericentre location given in Cartesian coordinates by  ${\bf R}_0 = (R_p, 0, 0)$.
Particles with $|{\bf R}|> R_p$ are reflected according to
\begin{eqnarray}
{\bf R} = \left( X', Y', Z' \right)  \rightarrow  \left(    2R_p - X', Y', Z'   \right)\\
\end{eqnarray}
By doing this, the particles then form a hemisphere   with  $ ,\sqrt{'X'^2+Y'^2} < R_p.$
The total mass of this is taken to be $0.25m.$
Most of these particles    will be on weakly  bound orbits  when they are given
the  pericentre velocity appropriate to  to a  zero energy orbit with pericentre at   $\left(    R_p , 0  , 0 \right)$  in the Cartesian system.
By following this procedure we omit consideration of the  $\sim 50 \%$ of the disrupting  star  that will be  unbound.
However,  this  does not intersect the disc and so it does  not play a significant role in our study. In addition the number of bound particles
that get transferred to the disc  from the stream  is increased.
The initial velocity of a particle  in the star is specified  in the spherical polar coordinate system  to be
${\bf v}_0 = \left(  0, 0  ,v_{\phi,0}\right), $
where
\begin{eqnarray}
v_{\phi,0}=\sqrt{{2GM\over R_p}}\sqrt{1+3{R_g\over R_p}}\ ,  {\rm with}\hspace{2mm}
R_g={{\mathrm{G} M} \over c^2},   R_g  {\rm \hspace{2mm} being \hspace{2mm} the \hspace{2mm}  gravitational \hspace{2mm}   radius }\nonumber \\
\nonumber
\end{eqnarray}
and  $c$ is the speed of light. Thus, neglecting  the effect of black hole  rotation which comes in at a higher order in $1/c,$ each particle is given the pericentre velocity  appropriate to a zero energy orbit
passing through pericentre. Being for the most part weakly bound they eventually return to the vicinity of pericentre in the form of a stream that persists till
arbitrarily large times. We remark that the initial configuration of the star is not in hydroststic equilibrium
and so pressure forces might be expected to produce some artificial expansion. However, the initial 
ratio of sound speed to orbital velocity
is $\sim 2\times 10^{-3}$ which is very  small 
Thus we anticipate that the effects of pressure imbalance to be  small until the stream first intersects itself. Tests we performed showed that until this stage, the motion 
of the stream was to a good approximation ballistic. At later times when the stream impacts disc material at larger radii $R/R_p  \sim 5,$ the  ratio of the width of the stream  to the local radius
 is  $\sim 0.1$ indicating the operation of some viscous spreading.  

\subsection{Equation of state }
\subsubsection{Disc particles}
For the disc, we adopt a locally isothermal equation of state  for which the locally isothermal sound speed
is given by $c_s = \delta |{\bf  v}_\perp|.$
Here, $\delta=H/R$ is the disc  semi-thickness  with $H$  being the disc scale height and we recall that $R=|{\bf  R}|$ is
the distance to the black hole. The
component of the  velocity vector of a particle that is perpendicular to ${\bf R}$
is ${\bf  v}_\perp$. In order to  determine ${\bf v}_{\perp}$ we set
\begin{equation}
 {\bf  v}_\perp={\bmth  \omega} \times {\bf   R} \ , \hspace{2mm} {\rm with}  \hspace{2mm}
 {\bmth  \omega}=  {\bf R\times v }  /{R^2}, \hspace{2mm} {\rm  where}
 \end{equation}
${\bf  v}$ is the velocity vector  of the particle and ${\bmth  \omega}$  its  angular velocity vector.
The direction of this is chosen so as to specify the required disc orientation.
The disc aspect ratio is chosen to be $\delta = 0.1$ for all simulations.

\subsubsection{Stream particles}

For the stream particles, we adopt an isothermal equation of state with a constant temperature of $T=10^6\ \mathrm{K}$.
The particles originating from the 'star' are evolved separately from the relaxed disc until just before the first particles return to periastron.
After this stage all the particles are allowed to interact. In so doing stream particles are converted  into disc particles.
The criterion we adopted for specifying when this first occurred for a particular particle
 was that the ratio of  binding energy to potential energy, calculated neglecting pressure and viscosity,
should have become $< 1/3.$ At that point the equation of state then switches from that for the stream to that for the disc,
provided the sound speed is larger
in the latter case. We remark that the small value of the ratio of the initial sound speed to orbital speed is found to result in the disc temperature  always exceeding the initial stream  temperature, accordingly heating occurs
 when a particle originating in the stream becomes tagged as a disc particle and then only the equation of state changes.

The total number of particles  involved in   the
simulations presented here is $4\times10^5$ with $50\%$ of these originating in the stream  and $50\%$ in the disc.
They have been checked by performing simulations with the particle number reduced by a factor $4$ which gave very similar results apart from in the very
central regions  with $R/R_p = r_p < 0.25$  where there are too few particles in the low resolution runs.

\subsection{A comparison of the disc inclination angle obtained from SPH simulations with semi-analytic results}

\begin{figure}
\begin{center}
\vspace{10cm}\includegraphics{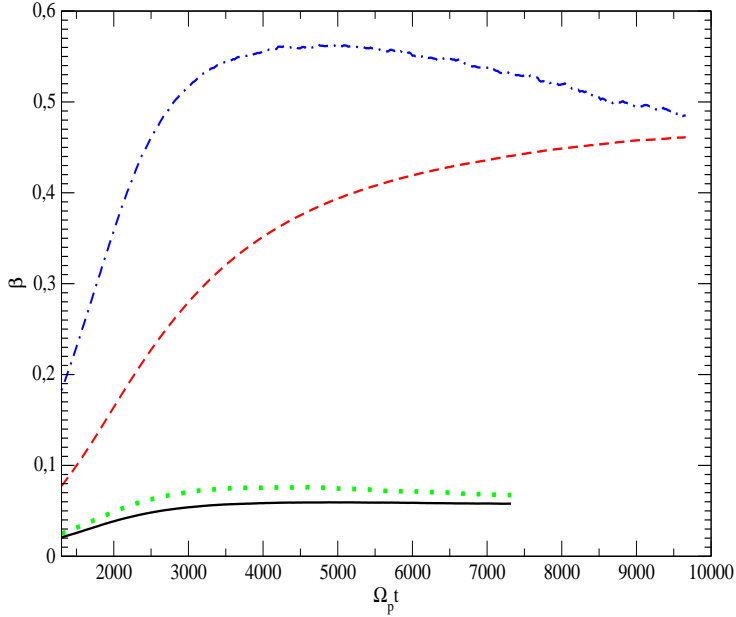}
\end{center}
\vspace{-0cm}
\caption{The dependence of the  inclination angle $\beta ,$ in radians,  at the stream impact position
 on time. See
the text for a description of particular curves.}
\label{FigSPH1}
\vspace{-0.0cm}
\end{figure}

\begin{figure}
\begin{center}
\vspace{12cm}\includegraphics{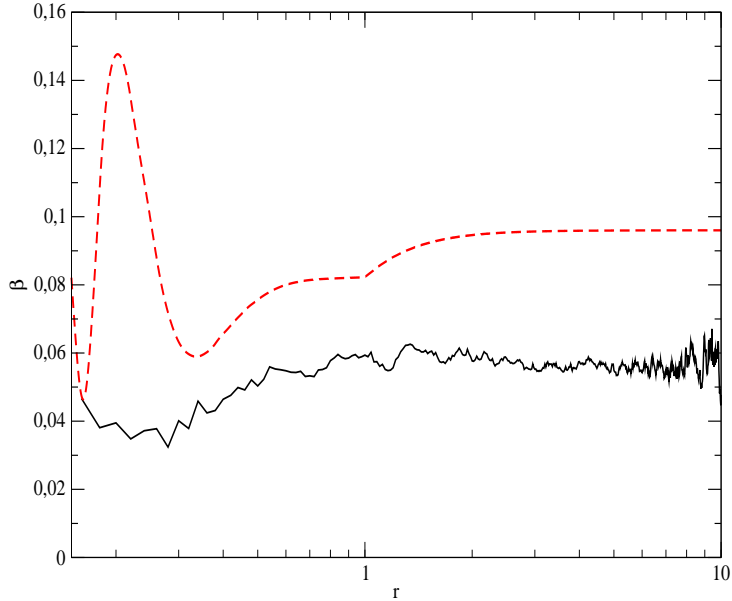}
\end{center}
\vspace{-0cm}
\caption{The inclination angle $\beta ,$ in radians,  shown as a function of radial distance $r_p$  for the case with $a=1$ at
time $t=3500\Omega^{-1}_p$.  The solid curve is from an SPH simulation, while {the dashed one
is } obtained by solution of (\ref{eq12}), see the text for details.}
\label{FigSPH2}
\vspace{-0.0cm}
\end{figure}

In order to make a comparison between the semi-analytic approach developed above and the SPH simulations
we use the  surface density $\Sigma $  and  the mass flux in the stream  obtained from simulations as input
background state variables to be used in  equation (\ref{eq12}).
 We determine a typical value of viscosity parameter
$\alpha $ by comparing  the dependence of the mass of the  accretion disc as a function of time obtained from SPH simulations of
a free accretion disc without the  presence of  the stream with  an analytic model based
on solution of the surface density evolution equation  which incorporates  an assumed  value of $\alpha ,$
 the latter quantity being chosen to provide the best match.
 The analytic model is the same as described  in \citet{Iva2015}, with the adjustment that the kinematic viscosity is taken to be  $\propto r_p^{1/2}$. This procedure
gives typical  values of $\alpha \approx 0.1$, which is employed in our solution of equation (\ref{eq12}) that yields the disc inclination angle.
\footnote{ We recall that the  disc inclination angle $\beta$ is defined as the angle
between the direction of the Z-axis of our (X,Y,Z) Cartesian coordinate system, which
coincides  with the direction of the  black hole rotation, and the  unit vector
perpendicular to the plane of a disc ring at a particular radius $r$.
Also, we do not show  behaviour at times prior to the
begining of stream-disc interaction, that is before the first stellar material returns to periastron.} 

We consider two cases, both having $\delta=0.1$, but different values of rotational parameter, namely
$a=1$ and $0.1$.
 The results of the  comparison are shown in Figures \ref{FigSPH1} and \ref{FigSPH2}. In Fig.  \ref{FigSPH1}
we show the dependences of  the disc inclination angle at the stream impact location  on time
Solid and dashed curves represent SPH results for $a=1$ and $0.1$, while dotted and
 dot-dashed curves are their respective counterparts obtained by solution of equation (\ref{eq12})\footnote{For the chosen values of
$\alpha$, $\delta $ and $a,$ the  parameters $\eta $
and $\eta_{rel}$ are approximately $0.6$ and $0.9$ for the $a=1$ case, and $0.1$, $0.4$ for the $a=0.1$ case.}
Note that all curves have been averaged over 100
data points
corresponding to the time spans $10^3\Omega_p^{-1}$ and $2\cdot 10^3\Omega_p^{-1}$ for $a=1$ and $0.1$,
respectively,
to remove numerical noise.
\begin{figure}
\begin{center}
\vspace{9cm}
\includegraphics{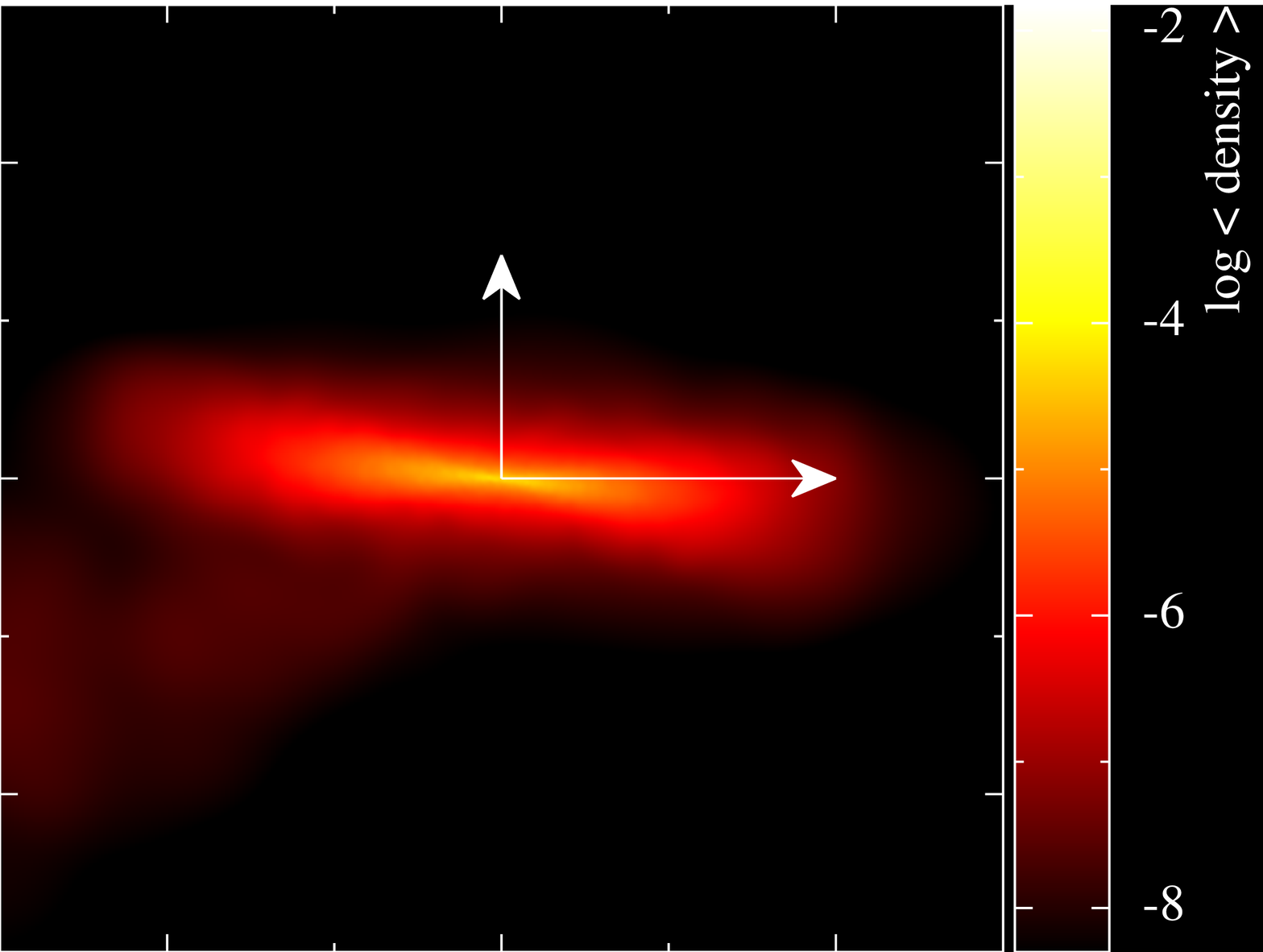}
\&
\includegraphics{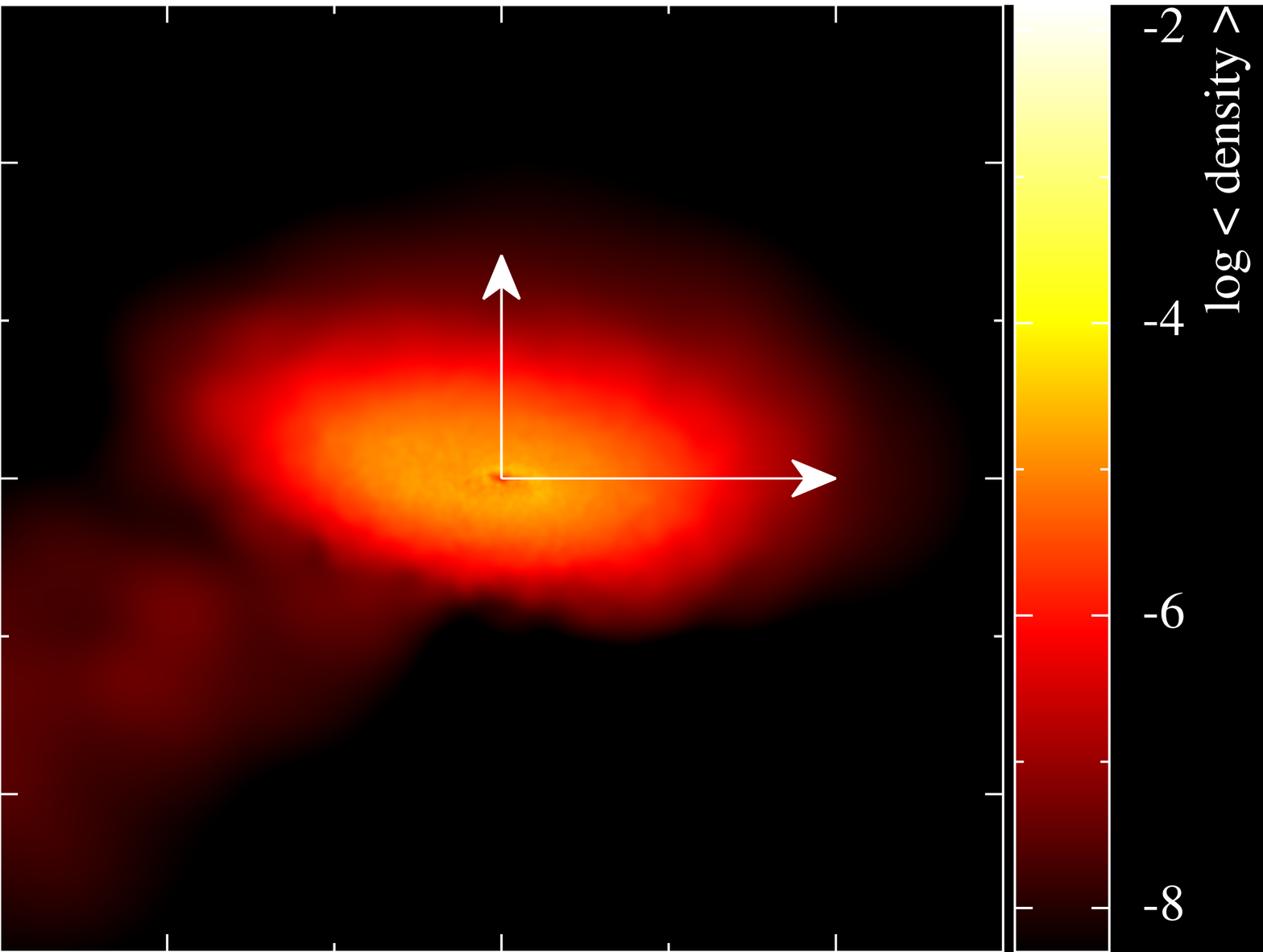}
\end{center}
\caption{The disc and stream seen in projection at $t= 8000\Omega_p^{-1}.$  The case with $a=1$ is illustrated in the left hand panel
and the case with $a=0.01$ in the right hand panel.
In these plots the long  vertical axis is perpendicular to the line of sight and  points in the direction of the black hole angular momentum.  The orthogonal axis shown
is the $X$ axis as defined in the text (see Section 2).  The tick marks on the axes of the plots are separated by $10$ dimensionless distance units.
The colour scale indicates $\log\langle \rho\rangle ,$  with  $\langle \rho \rangle $  being the projected density.}
\label{FigSPH2a}
\vspace{-0.5cm}
\end{figure}

We see that the analytic and SPH approaches are in agreement in finding that the case with $a=1$ becomes quite closely aligned whereas
 the case with $a=0.1$ maintains significant misalignment. Thus we can expect that for discs with $\delta \sim 0.1$ and other parameters appropriate to
 the TDE we consider, significant alignments are to  be expected for some systems.
 One can also  see from  Fig. \ref{FigSPH1}  that
there is a good agreement between the semi-analytic curves and those derived from SPH simulations corresponding to the case $a=1.$  The  typical deviation  is of
order of 20 per cent. When $a=0.1,$ there is a factor of $1.5-2$ disagreement for  times  ${ <  \approx 10^{4}\Omega_p^{-1}}$,
which can perhaps be attributed to a slower relaxation of the twisted disc to a quasi-stationary configuration in
this case. This is not unexpected as the time scale associated with attaining  alignment in the absence of the stream is expected to be longer for smaller values of the rotation parameter $a.$
 The disagreement becomes, however, quite small at later times with typical difference of the order of 5 per cent.
To illustrate the appearance of the disc we show  projections of the density distribution at $t=8000\Omega_p^{-1}$ for $a=1.0$ and $a=0.1$ in Fig. \ref{FigSPH2a}. It will be seen that 
in agreement with the above discussion, the case with $a= 1.0$ is almost aligned with the black hole equatorial plane whereas there
is significantly greater misalignment for $a=0.1.$



In Fig. \ref{FigSPH2} we show the  inclination angles as a function of radiius at  the
time $t=3500\Omega_p^{-1}$. Solid curve and dashed curves illustrate results respectively obtained using  the SPH method
and by solving  equation (\ref{eq12}) for $a=1$.
One can see that the disc inclinations change at approximately the same radial scale. {At radii $r_p > 1$ the
inclination angle stays approximately constant with its value being larger in the analytic approach
by about  $30-40\%.$ \footnote{Let us note that numerical grid based  magnetohydrodynamic calculations also
give typical values of the disc  tilt smaller than semi-analytic ones by factor of two, see Zhuravlev et al 2014.}.
This may be linked to the behaviour at small radii. For
$\sim 0.4 < r_p  < \sim 1$ the inclination angle decreases torwards black hole in both approaches. However,
there is a qualitative difference in the behaviour of disc's tilt at smaller $r_p$. While in the analytic
approach the inclination angle oscillates with a rather large amplitude, the numerical simulations give a
much more moderate evolution. This difference can be attributed to an insufficient number of SPH
particles to fully resolve the  oscillations of  the disc  tilt.  In our case, $N=4\cdot 10^5.$
We remark that  a specially dedicated study shows that the number of particles required to achieve this should be an  order of magnitude
larger (see Nealon et al. 2015).

To summarise there is full qualitative agreement between the two  approaches apart from the issue of the disc
tilt behaviour at small radii, which is not important for our purposes. For the significant  quantities and/or values of
parameters there is  quite good quantitative agreement. This validates the use of our analytic approach for much
more realistic backgrounds evolving over much longer times $\sim 10^5$ orbital periods at periastron
  provided by our grid based 1D  finite difference numerical simulations ( see
 Section \ref{Grid} below) }.  It is impracticable to carry out SPH simulations for  such long times.



\section{Grid based calculations of thermally unstable discs}\label{Grid}

We have performed grid based simulations of an accretion disc
supplied by a stream produced by a tidally disrupted star.
The expected accretion rate is sufficiently large that the Eddington limit is exceeded,
radiation pressure becomes important and the disc undergoes thermal instability.
This instability produces transitions between high and low states that in some circumstances
can move as propagating fronts. When the accretion rate is approximately steady in the mean  cyclic behaviour may occur
in which there is alternation between  high and low states.

The formalism is similar in concept to that of \citet{Szu1998}.
who considered the evolution of thermal instabilities in accretion discs for which radiation
pressure plays a significant role..
However, it is adapted to conditions around a $10^6M_{\odot}$ black hole
rather a $10M_{\odot}$ black hole that they studied.  In addition radiative transfer in the radial direction  and mass input from
an accretion stream  produced by tidal disruption is considered.

Furthermore,  our simulations have been performed with NIRVANA
which is a three-dimensional  MHD code that has been
described in e.g. \citet{Zie1997}. For the grid based calculations  performed  here
the magnetic field is set to zero and  the code is restricted to operate in one dimension.
For convenience we retain spherical polar coordinates restricted to the plane $\theta=\pi/2$
and assume the system is independent of the azimuthal coordinate.
The sole independent spatial coordinate,  being the radius, $R,$  can then be regarded  as a cylindrical coordinate.
Note that we solve the governing equations in Eulerian form and so our procedure differs
markedly from the one dimensional Lagrangian approach implemented by \citet{Szu1998}.

 \section*{Basic equations }

The basic equations are those of
 mass,  momentum  and energy conservation appropriate to a mixture of gas and radiation.
In a non rotating frame with origin at the location of the black hole, these
  take the form
\begin{eqnarray}
\frac{\partial \Sigma}{\partial t} + \del \bcdot (\Sigma \bb{v})  &=&  S_m \,\label{contg} , \\
\Sigma \left(\frac{\partial \bb{v}}{\partial t} +  \bb{v} \bcdot \del \bb{v} \right)   &=&  -  \del  {\Pi}
   -\Sigma \nabla \Phi  +{\bf f}_v \,\label{mog} ,\\
   \frac{\partial { E} }{\partial t}  +  \del {\bcdot} (E \bb{v}) &=&  -{\Pi} \del {\bcdot }({\bf v})    + \epsilon_v  -   2 F_{+}   - \nabla \cdot (2H F{\hat R})+ S_{E},  \label{engg} ,
\label{basic_eq}
\end{eqnarray}
Here,  ${\bf v} = V_R {\bf \hat{R}}$ is the velocity, with ${\bf \hat{R}}$ being  the unit vector in the radial direction,
${\Pi}$ is the vertically integrated   pressure,
 and $E$ is  the vertically integrated internal energy per unit volume.
 The viscous force per unit area is ${\bf f}_v,$   the rate of energy input per unit area due to viscous dissipation is $\epsilon_v$
 and the radiation flux per unit area leaving one side of the disc is $F_+.$ The factor of two multiplying this quantity accounts
 for the two sides of the disc. The radiative flux  in the radial direction is
\begin{eqnarray}
F= - \frac{c}{\kappa \rho}\frac{ \partial  (a_RT^4/3) }{\partial R}.
\end{eqnarray}
The opacity is $\kappa$ and $a_R$ is the  Stefan Boltzmann constant.
The mass input rate per unit area from the stream is $S_m,$ which is discussed further in Section \ref{Inputstream} below,
and the rate of excess thermal energy input per unit area associated
with this is $S_E.$
For these studies we adopt the Paczynski Wiita  potential such that  $\Phi = -GM/(R-2R_G).$
 Here, as elsewhere,  the self-gravity of the gas is neglected.

\subsection{Equation of state }

The two state variables we use to characterise the mixture of gas and radiation are
$\Sigma$ and $E.$ These are related to the mid plane density and internal energy density
through $\Sigma =2H\rho$ and $E=2HU$ respectively. where $H$ is an effective semi-thickness.
We make the assumption that this is the same for both quantities and also that $\Pi =2HP,$
 where $P$ is the mid plane pressure. Then from the expressions for  $P$ and $U$ in terms of $\rho$ and $T$
 given  by
\begin {equation}
P = ({\cal R}/\mu)\rho T+a_RT^4/3 \hspace{2mm}  {\rm and} \hspace{2mm} U= (3{\cal R}/2\mu)\rho T+a_RT^4,
\end{equation}
on the disc mid plane we obtain
\begin{equation}
\Pi=  ({\cal R}/\mu)\Sigma  T+2Ha_RT^4/3  \hspace{2mm} {\rm and} \hspace{2mm} E= (3{\cal R}/2\mu)\Sigma T+2Ha_RT^4.\label{EOS}
\end{equation}
Here the mean molecular weight is $\mu$ and  ${\cal R}$  is the gas constant.
Once $H$ is related to  $E$ and $\Sigma$ all of the local state variables can be found.
For example the temperature is found by solving the quartic equation obtained from  the second expression in (\ref{EOS}).
We adopted the expression
\begin{equation}
   H=\sqrt \frac{ 2\sqrt{2}E}{
          3\Sigma  \Omega^2}
\label{thick1}
\end{equation}
for the semi-thickness  $H.$
Regarding this, we recall that from vertical hydrostatic we get the conventional estimate
\begin{equation}
H= \sqrt{2 P/(\rho\Omega^2)}.\label{thick}
\end{equation}
 Noting that in the gas pressure dominated limit we have
$P/\rho= 2E/(3\Sigma)$ and in the radiation dominated limit we have $P/\rho= E/(3\Sigma),$
the value of $H$ obtained from (\ref{thick1}) differs from (\ref{thick})  by at  most $2^{1/4}.$
As this is within the uncertainties inherent in  carrying out the vertically averaging procedure, we
 adopt   (\ref{thick1}). For the calculations reported here, the mass fraction in hydrogen was taken to be $0.7$
 and $\mu =0.615.$

\subsection{Radiative cooling and viscosity}

The emergent radiation flux is given by
\begin{equation}
     F_+ =\frac{2a_RcT^4}
       {(3\kappa \Sigma +4/3)}.
       \label{Flux}
\end{equation}
 Equation (\ref{Flux}) is obtained by
writing $  F_+ = a_RcT^4_{{\rm eff}}/4,$ where $T_{{\rm eff}}$ is the effective temperature.
Using the Eddington approximation to relate  $T$ and $T_{{\rm eff}}$ then gives (\ref{Flux}).
We remark that $\kappa\Sigma/2$ is the optical depth of the mid plane.
For the model considered here we assume that the opacity is due to electron scattering
and accordingly is constant.

Viscosity is incorporated through adopting the standard $\alpha$ parameterisation of \citet{Sha1973}. Using this the kinematic viscosity is given by
\begin{equation}
\nu= \alpha P/( R\rho | d\Omega/dR |) \hspace{2mm}{\rm leading \hspace{1mm} to \hspace{2mm}} \langle \nu\rangle = \alpha \Pi/( R\Sigma |d\Omega/dR |),
\end{equation}
where $\langle \nu \rangle $ is the density weighted  vertically averaged viscosity.

When the disc becomes thermally unstable, it can becomes very thick, with $H/R$ driven to values exceeding  unity.
Our formalism then becomes inappropriate. It is also likely that some of the available energy goes into driving an outflow rather
than mid plane heating. Accordingly we have limited the  heat production rate when the disc becomes thick.
For the runs  considered we applied a reduction factor
$(1-(H/R))^2$  to the heating rate.   Although this quenching is ad hoc it limits  the growth of $H/R$
of the disc region modelled, but otherwise does not  affect the qualitative form of the results.

\subsection{Computational domain and boundary conditions}

Simulations were performed over the radial domain $[ R_{in}, R_{out}] $
with  $R_{in} =  [3.0388\times 10^{12} cm$ and $R_{out} = 1.484012\times 10^{13} cm. $
We have employed $N_g =768$ equally spaced grid points and checked convergence
using twice as many ($N_g = 1536).$

At both radial boundaries we employ a limited outflow condition.
This is the same as the standard outflow condition but  with the additional
feature that the magnitude of the outflow velocity is limited
to be less than or equal to $3\langle \nu \rangle/(2R_{out}).$
These velocities are characteristic of the inflow velocity driven by viscosity
and  will remain small if the disc remains thin. However, they can become moderately large
without causing a major pile up of mass
when the disc becomes thick,  if $\alpha$ is not too small. Thus advective transport can arise
when the disc becomes thermally unstable.

\begin{table}
\begin{tabular}{lll}
 Model & $\alpha$ & $f_{st}$ \\
 A  & 0.3 & 1.0 \\
 B  & 0.1 & 1.0  \\
\end{tabular}
\caption{Parameters of models for  which results are described in the text.
\label{table1}}
\end{table}

\subsection{Input from the stream}\label{Inputstream}

For a star of $1M_{\odot},$ the pericentre distance  for a penetration factor $B_p$
is given by
$R_p = 7\times 10^{12}(M/(10^6M_{\odot})^{1/3}/B_p$ (see Section \ref{Char}).
For our simulations we take $M = 10^6M_{\odot}$
and $B_p=  14/9.$  Then the pericentre distance is $4.5\times 10^{12} cm$ and the minimum return time is
$P_{min} = 3.5\times10^6 B_p^{-3} s.     =  9.30\times 10^5 s.  $

We adopt the following simplified prescription for the accretion rate ${\dot M_S}$  from the stream generated
by the tidally disrupted star.
Setting $t=0$ to be the time of pericentre  passage, we take ${\dot M_S} =0$ for $t < P_{min}.$
For $t > P_{min},$ we set
${\dot  M_{s}}  =  7.17\times 10^{26}(t/P_{min})^{-5/3} gm s^{-1}.$
Thus we assume a tail off $ \propto t^{-5/3}$  (Lodato et al 2009 and references therein) and a total mass to be  accreted of $0.5M_{\odot}.$ This mass  accretion rate is input uniformly over $8$ grid cells centred on the circularisation radius $R_S = 2R_p$ at the lowest resolution. For higher resolutions the number of grid cells  used is proportional to the resolution. This procedure determines $S_m.$  We remark that this simplified model assumes that the stream has a high enough density that it is able to penetrate any intervening disc material in order to reach
the circularization radius, avoiding prior significant angular momentum exchange. This scenario may require  modification when the disc becomes very thick at late times when
the accretion rate is small.
However, for simplicity  we adopt it throughout.   

When mass from the stream enters the disc energy is dissipated.  Assuming the plane of the stream is only slightly inclined
to that of the disc, the kinetic  energy  per unit mass
associated with radial motion available to be  dissipated is $GM/(2R_S).$ Depending on details of the circularisation process
a part of this is radiated away directly and a part is converted to excess internal energy of the disc. This energy is input along
with and in the same way as the mass input in this way $S_E$ is determined.
 For simplicity we have assume this
fraction to be $ f_{st}= 50\%$ for the simulations presented here.
However, we have also run cases with this input reduced by more than an order of magnitude.
We have found that this does not change the qualitative form of the results significantly
as the internal energy provided by dissipation of stream kinetic energy ultimately never dominates
that arising from viscous dissipation throughout the disc.

\subsection{Initial disc}

The steam commences to  input mass at $t= P_{min}$ into an  initial disc.
This was specified to have  a low mass of $0.011M_{\odot}$
as compared to the total to be input from the stream.
The state variables were chosen such that $\beta = 0.5$ was constant
with  $\rho \propto R^{-3}$  for $R  < 7.572\times 10^{13}$ and   $\rho \propto R^{-1}$
for $R < 7.572\times 10^{13}.$ With this choice $\Sigma  \propto R^{-2}$  for $R < 7.572\times 10^{13}$ and   $\Sigma \propto R^{1/3}$
$R >7.572\times 10^{13}.$   We remark that  as after a short time  the simulation is dominated by the mass  input from
the stream,  which  occurs at a rate implying that the Eddington limit is exceeded (see Section \ref{edd} below),  results are not expected to be affected by the choice of initial disc structure.
Profiles are rapidly modified by outward propagating transition fronts. Thus reducing the initial
value of $\beta$ by a factor of two has no significant effect.

\subsection{Simulation  of a flat disc }
\begin{figure}
\begin{center}
\vspace{20cm}\hspace{-18cm}\includegraphics{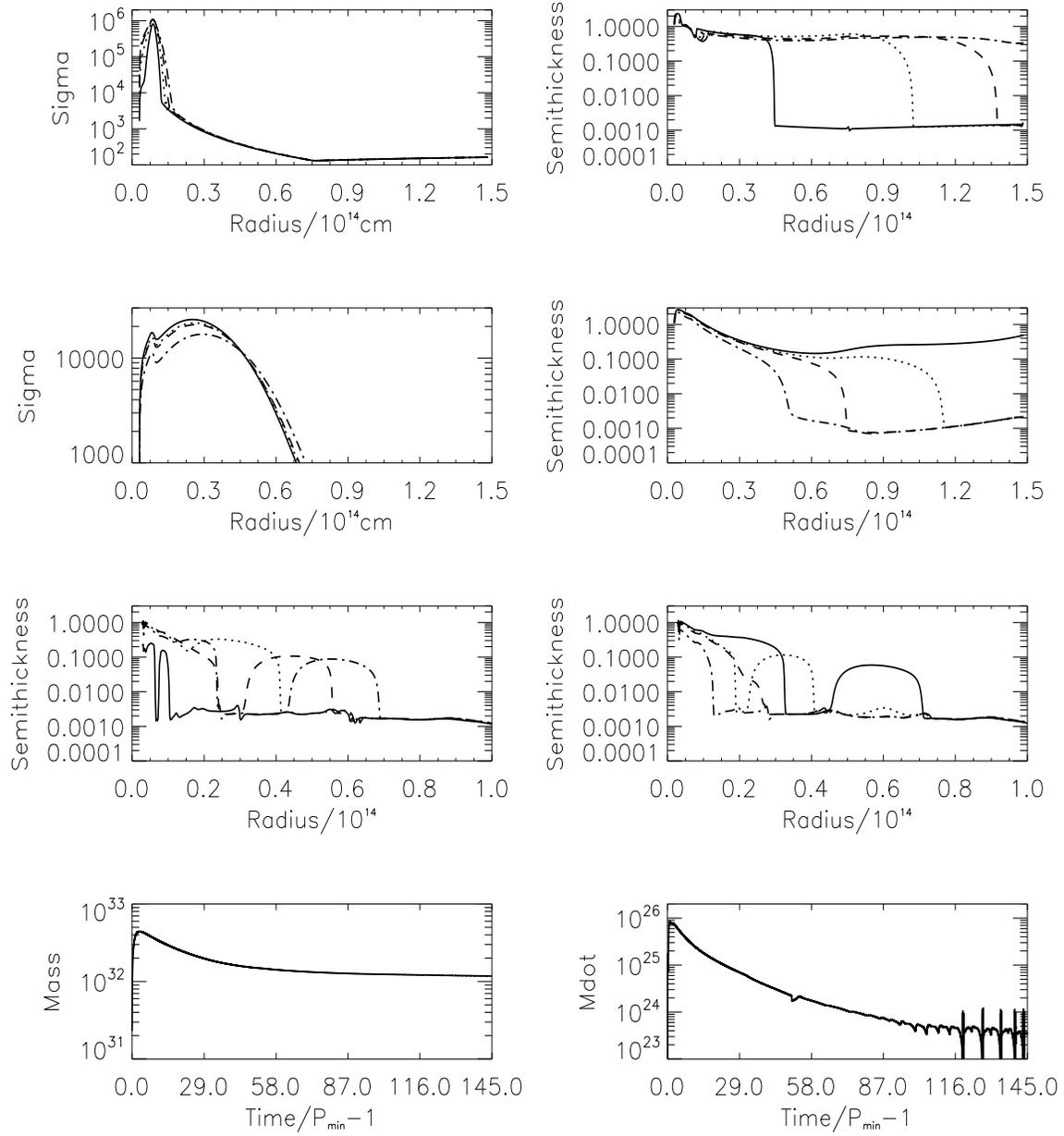}
\end{center}
\vspace{-1cm}
\caption{
Simulation results for model $A$ with $\alpha =0.3$ are shown. the uppermost panels show the evolution of the first  outwardly
propagating transition front. The left hand panel shows the surface density and the right hand panel the semi-thickness.
 Their functional forms  are given  at times expressed in the form $(t-P_{min})/P_{min}$
 of,  $0.216,$ (solid curve)  $0.432,$ (dotted curve)
 $0.649$ (dashed curve)  and $0.865$ (dot-dashed curve).
The second row of  panels show the evolution of  the first inwardly
propagating transition front. The left hand panel shows the surface density and the right hand panel the semi-thickness.
Their functional forms are given at times expressed in the form  $(t-P_{min})/P_{min}$
 of,  $37.83,$ (solid curve)  $40.00,$ (dotted curve)
 $41.08$ (dashed curve)  and $47.56$ (dot-dashed curve).
The third  row of  panels show the evolution of a later  outburst.
 The left hand panel shows the semi-thickness during the heating phase
 and the right hand panel the semi-thickness during the cooling phase.
Their functional forms are given at times expressed in the form  $(t-P_{min})/P_{min}$
 of,  $118.91,$ (solid curve)  $120.10,$ (dotted curve)
 $121.29$ (dashed curve)  and $122.53$ (dot-dashed curve) for the left hand panel and
of,  $122.96,$ (solid curve)  $123.72,$ (dotted curve)
 $124.91$  (dashed curve)  and $126.15$ (dot-dashed curve) for the right  hand panel .
The lowermost left panel shows the mass in the disc in $gm$  as a function of time. The lowermost right panel shows the
accretion rate into the central part of the disc in $gm s^{-1}$ as a function of time.}

\label{FigG1}
\vspace{-0.5cm}
\end{figure}

Simulation results for model $A$ with $\alpha =0.3$ are illustrated in Fig. \ref{FigG1}. the uppermost panels show the evolution of the first  outwardly
propagating transition front. The forms  of the surface density and the semi-thickness are plotted.  A front  is seen to have formed after a time given by
 $(t-P_{min})/P_{min} = 0.216$  and reaches the outer boundary  after a time given by $(t-P_{min})/P_{min} = 0.865.$
 At this stage the disc attains  a high  state with $H/R  \sim 1$ in the inner and outer parts, being somewhat smaller in the central
 regions.    After a time given by  $(t-P_{min})/P_{min} = 37.83$ an inwardly propagating front is seen in the outer parts of the disc.
Its evolution is shown in the second row of  panels of Fig. \ref{FigG1}. After a time  approximately given by
$(t-P_{min})/P_{min} = 47,$ this stalls at a radius $R\sim  5\times 10^{13}cm$ and the evolution enters a quasi steady phase
with the outer disc in a low state.
 The total mass content of the disc as well and the accretion rate through the boundary are
 shown in the lowermost panels of Fig. \ref{FigG1}.  It will be seen that during the time the disc is in a high
 state between these upward and downward transitions,
 around $0.1M_{\odot}$ is accreted through the centre by means of a strong advective flow.
 At later times a series of cycles in which parts of the disc alternate between high and low states.
 The presence of these can be seen through the behaviour of the mass accretion rate into the central regions at late times
 where several distinct oscillations can be seen. Before this behaviour is noticeable
there are several outbursts for which the whole disc is again in a high state.  At increasingly late times the outbursts become progressively
 more confined in the central parts of the disc with the outer parts remaining in a low state. This is a naturally expected outcome
 as the mass input  rate from the stream declines towards zero.
The third row of  panels illustrates the evolution of the outburst that occurs
for $ 18.91  126.15 < (t-P_{min})/P_{min} < 128.15.$
During this outburst the accretion rate into the centre is affected and the outer parts of the
disc remain in a low state throughout. Also during the heating phase the disc there are two regions in a high state
separated by a region in a low state.

\begin{figure}
\begin{center}
\vspace{20cm}\hspace{-18cm}\includegraphics{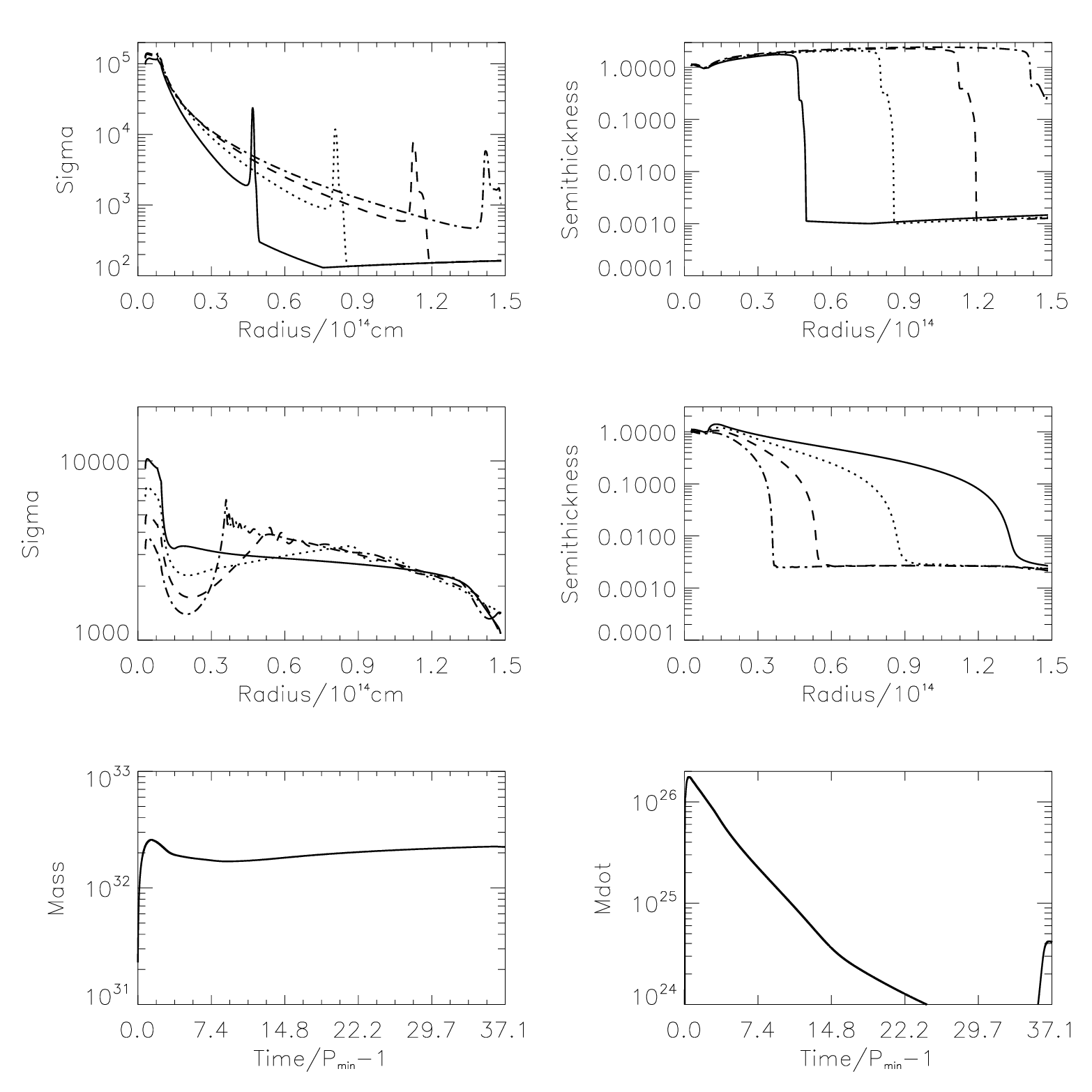}
\end{center}
\vspace{-1cm}
\caption{
Simulation results for model $B$ with $\alpha =0.1$ are shown. the uppermost panels show the evolution of the first  outwardly
propagating transition front. The left hand panel shows the surface density and the right hand panel the semi-thickness.
 Their functional forms  are given  at times expressed in the form $(t-P_{min})/P_{min}$
 of,  $0.649,$ (solid curve)  $1.297,$ (dotted curve)
 $1.946$ (dashed curve)  and $2.594$ (dot-dashed curve).
The second row of  panels show the evolution of  the first inwardly
propagating transition front. The left hand panel shows the surface density and the right hand panel the semi-thickness.
Their functional forms are given at times expressed in the form  $(t-P_{min})/P_{min}$
 of,  $29.186,$ (solid curve)  $34.051,$ (dotted curve)
 $38.915$ (dashed curve)  and $44.320$ (dot-dashed curve).
The third  row of  panels show the evolution of a later  outburst.
 The left hand panel shows the semi-thickness during the heating phase
 and the right hand panel the semi-thickness during the cooling phase.
Their functional forms are given at times expressed in the form  $(t-P_{min})/P_{min}$
 of,  $45.131,$ (solid curve)  $45.779,$ (dotted curve)
 $46.158$ (dashed curve)  and $46.482$ (dot-dashed curve) for the left hand panel and
of,  $47.563,$ (solid curve)  $49.725,$ (dotted curve)
 $51.887$  (dashed curve)  and $54.319$ (dot-dashed curve) for the right  hand panel .
The lowermost left panel shows the mass in the disc in $gm$  as a function of time. The lowermost right panel shows the
accretion rate into the central part of the disc in $gm s^{-1}$ as a function of time.}
\label{FigG2}
\vspace{-0.5cm}
\end{figure}

Simulation results for model $B$ with $\alpha =0.1$ are illustrated in Fig. \ref{FigG2}. the uppermost panels again how the evolution of the first  outwardly
propagating transition front.  In this case the front  is seen to have formed after a time given by
 $(t-P_{min})/P_{min} = 0.649$  and reaches the outer boundary  after a time given by $(t-P_{min})/P_{min} = 2.594.$
It is accordingly about a factor of two slower than for model A. The entire disc again goes into a high state with $H/R \sim 1$.
However, its duration is some what shorter in this case.
 After a time given by  $(t-P_{min})/P_{min} = 29.186$ an inwardly propagating front can be  seen in the outer parts of the disc.
Its evolution is shown in the second row of  panels of Fig. \ref{FigG2}. After a time  given approximately  by
$(t-P_{min})/P_{min} = 44$ this stalls at a radius $R\sim  4\times 10^{13}cm$ and the evolution again  enters a quasi steady phase
with the outer disc in a low state.
The total mass content of the disc as well and the accretion rate through the boundary
 shown in the lowermost panels of Fig. \ref{FigG2} indicate that  when  the disc is in a high state between these transitions
 around $0.05M_{\odot}$ is accreted through the centre being about half of that found for model A.
As for model A, a sequence of cyclic heating and cooling events occurs. However, these are more confined to
the outer disc and tend to involve smaller radial scales, which decrease as time progresses.
The third row of  panels illustrates the evolution of the outburst that occurs
for $ 45.131  < (t-P_{min})/P_{min} < 54.319.$  In this case although the region of the disc closest to the outer boundary
remains in a low state, the heating commences in the outer part of the disc converting a section into a high state
with very little effect on the innermost regions. During the cooling phase, the high state outer region collapses while the innermost
high state region  retreats inwards. Continuation of this calculation, as for model A,  results
in outbursts that become progressively
 more confined in the central parts of the disc  as the accretion rate  reduces with the outer parts remaining in a low state
 but with small scale fluctuations.

\subsection{The evolution of the inclination angle  for background disc models obtained from the grid based simulations}\label{Gridmodels}

In order to find the evolution of twist and tilt angles as a function of  time we numerically  integrate equation (\ref{eq12}) using the grid based  models
described above to specify the  dependence of the  semi-thickness,  $\delta, $ and  surface density,  $\Sigma, $ on time, for  model  A  with  $\alpha=0.3$ and
model B  with $\alpha =  0.1$ (see table \ref{table1}).  Other disc parameters are fixed as indicated above.
Note that
the surface density enters equation (\ref{eq12}) implicitly through the quantities, $\sigma$, defined  through  equation (\ref{eq14}),  and $\xi=
\delta^2 \Sigma r^{1/2}.$ To find the mass flux in the stream we use equation (\ref{eq4}) when $t > P_{min}$
and assume that $\dot M_{s}=0$ at earlier times. The outer radius of the integration domain was fixed to be $r_{out}=18$, while the inner radius  was taken to be $r_{in}=0.25$
for cases where  the inclination angle exponentially grows for  $r < 1$ and  $r_{in}=0.04$ for  cases  where  it oscillates at small values of $r$. We consider a range of values
for the black hole  rotation parameter
$a$ and both prograde and retrograde rotation. Note that although the background models were obtained with this set to zero, they are not expected to have a significant
dependence on it. However, this is not the case for the disc inclination when the disc  suffers a misaligning perturbation.

\begin{figure}
\begin{center}
\vspace{12cm}\includegraphics{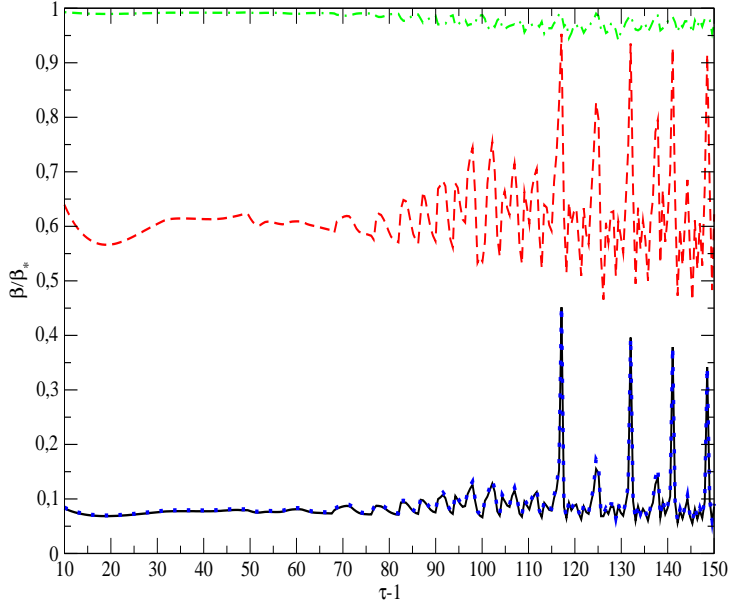}
\end{center}
\vspace{-0cm}
\caption{The value of inclination angle $\beta$ at the stream impact position $r=1$ as a function of time $\tau=t/P_{min}$ shown for
the grid based model with $\alpha=0.3$. See the text for a description of different particular curves.}
\label{Fig6}
\vspace{-0.0cm}
\end{figure}

\begin{figure}
\begin{center}
\vspace{12cm}\includegraphics{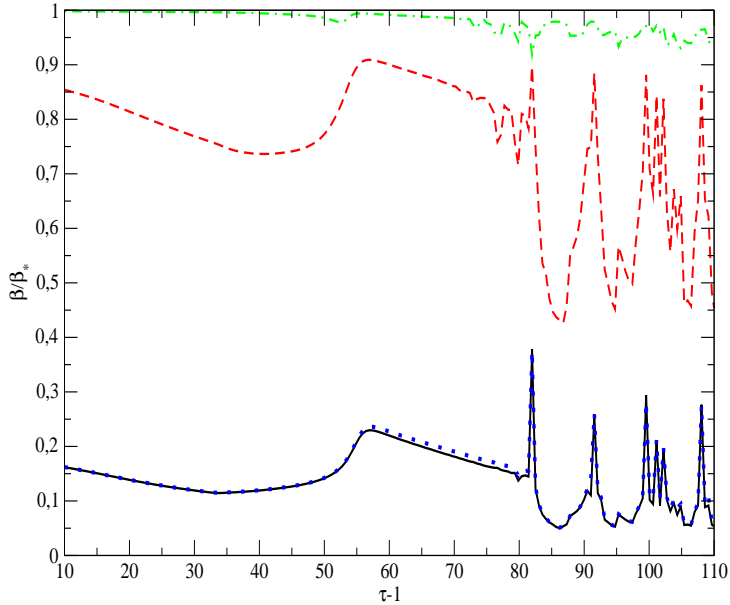}
\end{center}
\vspace{-0cm}
\caption{The value of inclination angle $\beta$ at the stream impact position $r=1$ as a function of time $\tau=t/P_{min}$ shown for
the grid based model, B,  with $\alpha=0.1$. See the text for a description of different particular curves.}
\label{Fig7}
\vspace{-0.0cm}
\end{figure}

We show the value of the inclination angle $\beta $ in units of the stream inclination $\beta_*$ as a function of time $\tau=t/P_{min}$ in Figures \ref{Fig6} and \ref{Fig7},
respectively, for models A and B.

In Figures  \ref{Fig6} and \ref{Fig7} solid, dashed and dot-dashed lines  illustrate  calculations for prograde rotation of
the black hole  with $a=1$, $0.1$ and
$0.01$, respectively. From these  results it is apparent  that smaller black hole rotations lead to larger disc inclinations, as expected.
In addition the time averaged  values of the inclination for $a=0.1$ and $a=1$ are similar in magnitude to  those found for the SPH simulations
of a disc with $\delta \sim 0.1$ throughout as discussed in Section \ref{SPH}.
The dotted curves illustrate the case of retrograde rotation with $a=-1.$  Since   results found for retrograde
rotation with a  smaller absolute value  of $a$ practically coincide with those obtained for  its  prograde counterpart, they are not shown.

The results illustrated in  Figs. \ref{Fig6} and \ref{Fig7}  indicate
 that values of the scaled  inclination are quite substantial for our models at all times,  being  of the order of $0.1$
for $a=1$ and  larger for smaller values of rotational parameter,
The high state super-Eddington regime of accretion corresponds
to $\tau = t/P_{min}  < \tau_{crit}\approx 80-100$.
 When $\tau > \tau_{crit}$ a transition to the low state occurs, but, since there is a continuing supply of mass from the stream  and radiation pressure is important,
the condition for the development of thermal instability will become satisfied with the result that the disc undergoes a sequence
of transitions between high and low states. The inclination angle changes quite dramatically in course of these transitions
being order of the maximal values at low states and dropping to values order $(0.05-0.1)\beta_*$ during the intermittent high states.

\begin{figure}
\begin{center}
\vspace{12cm}\includegraphics{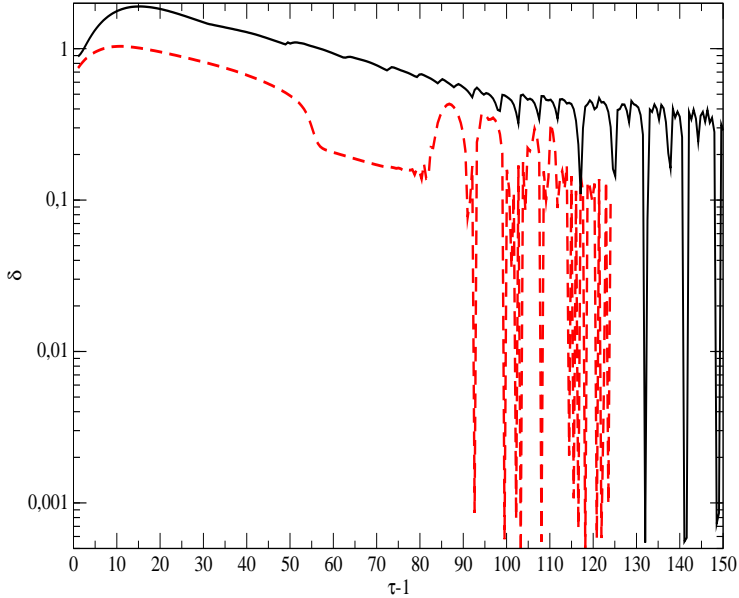}
\end{center}
\vspace{-0cm}
\caption{Dependencies of $\delta (R=0.5R_S)$ on time are shown. Solid and dashed lines correspond to models A and B, respectively.}
\label{Fig7n}
\vspace{-0.0cm}
\end{figure}

These sharp changes  in disc  inclination are related to sharp changes of the  aspect ratio $\delta $ during these transitions.
In order to illustrate this we   show the dependence of $\delta (R=0.5R_S)$ on time in Fig. \ref{Fig7n}. Solid and dashed lines represent
models A and B, respectively. One can see from this Figure that the inclination angle and disc semi-thickness experience strong variations
 at the same time.

\begin{figure}
\begin{center}
\vspace{12cm}\includegraphics{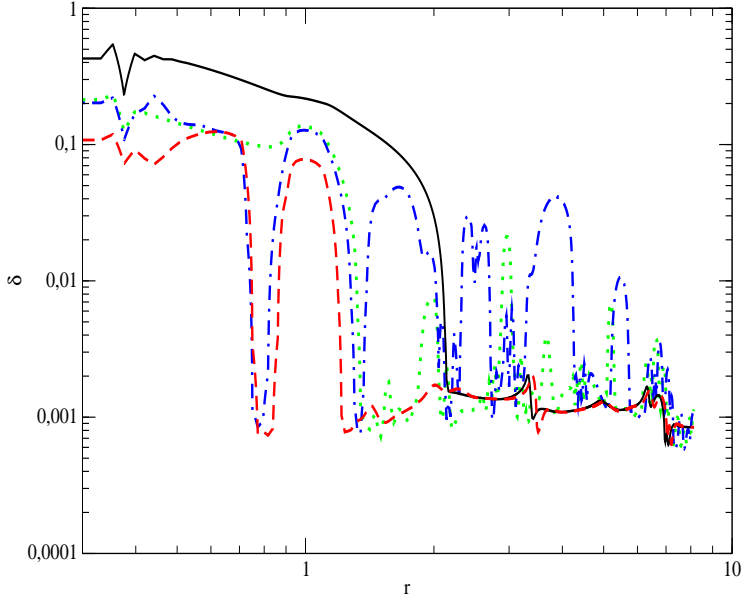}
\end{center}
\vspace{-0cm}
\caption{The dependence of the disc semi-thickness $\delta $ on radius $r$ for different disc  models at different
times, see the text for description of  the different  curves.}
\label{Fig8}
\vspace{-0.0cm}
\end{figure}

\begin{figure}
\begin{center}
\vspace{12cm}\includegraphics{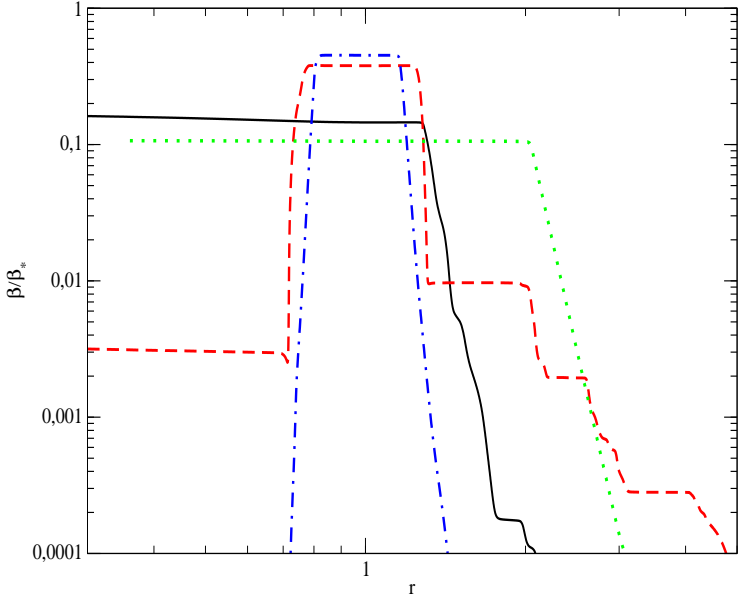}
\end{center}
\vspace{-0cm}
\caption{Same as Fig. \ref{Fig8}, but the form of the  inclination angle $\beta $ is shown, see the text for description of different lines.}
\label{Fig9}
\vspace{-0.0cm}
\end{figure}


We show the functional form of $\delta $ and $\beta $ with radius in models A and B immediately before
and after the first transition from a high to a  low state In Figs. \ref{Fig8} and \ref{Fig9}.   .
Solid and dashed curve are for model A at $\tau=115.5$ and $\tau = 117$
while the dotted and dot-dashed curves are for model B at $\tau=80.4$ and $\tau=82$.
Solid and dotted curves apply during the high
state, while dashed and dot dashed curves are for
the later occurring low state.
It is seen that there is a clear relationship between a sharp decrease of $\delta $
in the vicinity of $r\sim 1$ and a corresponding growth of inclination angle.
Note that this  is more prominent for  model A. Also note that even in the low state $\delta $ is much
larger that its 'low' value of  $10^{-3}$ very close to $r=1$. This is related to heating the disc by the stream.


\section{Analytic estimates for $\tau_{crit}$ and the possibility
of large values of the disc inclination during the transition to a 'low state' }\label{edd}

In our numerical work we  have considered disc models  for  only  a small range of parameters characterising them.
Therefore, it is important to obtain estimates of the transition time $\tau_{crit}$ and
maximal values of $\beta $ during these transitions, which can be applied to  models  characterized  by  a wider
range of parameters  such as  any value of $\alpha < 1.$

A crude estimate of the transition time $\tau_{crit},$ at which a major part of the disc first transitions from a high to low state,
can be obtained from the following simple
considerations (see also Shen $\&$ Matzner 2014 and references therein). During the high state  energy
dissipated through the action of  viscosity  per unit  time in a disc  annulus centred at
radius $R$ and of unit width, is mainly advected towards black hole.
It is accordingly  larger than the amount  of energy radiated away by this annulus.
The rate of dissipation  of energy per unit radial width, $\dot E_{diss}$,  can be related to the mass flux $\dot M$ through
 $\dot E_{diss}={3GM/( 2R^2)}\dot M$ \citep[e.g.][]{Abr1988}, while the rate of radiation of energy
from the annulus is equal to the flux, $F_{+}$, multiplied by
$4\pi R$ to take into account of both disc surfaces. In the advective stage we therefore
have
\begin{equation}
\dot E_{diss}={3GM\over 2R^2}\dot M > 4\pi RF_{+},
\label{en1}
\end{equation}
where we assume that  angular frequency of the disc material  is approximately Keplerian.

As  discussed above, in the optically thick limit we specify
 $F_{+}=2 c P_{r}/( \kappa \Sigma)$, where $\kappa $ is the Thomson opacity
and $P_{r}$ is radiation pressure.
We also use the vertical component of the  hydrostatic equilibrium equation
which  leads to $P={GM/( 2^{5/2}R^3)}\Sigma H$, where $P$ is the total pressure and we recall
that $H=\delta R$.

We  now assume that $P\approx P_{r}$ and express the mass flux in terms of the Eddington value $\dot M_{E}= {4\pi G M/( c \kappa) }$ as
$\dot M =\dot M_{E}\dot m$ which introduces a dimensionless accretion rate ${\dot m}.$
 We then substitute the resulting expression in equation (\ref{en1}) and use the expressions for radiation flux
and the total pressure  given above to obtain
\begin{equation}
\dot m > {\delta \over 3\sqrt{2}}{R\over R_g},
\label{en1n}
\end{equation}
in order that  advective cooling  overcomes  radiative cooling.
Now we assume that $R\sim R_S$, use the definition of
$B_S$ and equation (\ref{eq1}) to get
\begin{equation}
\dot m > 11\delta M_{6}^{-2/3}/B_S.
\label{en2}
\end{equation}
Note that this implies that a thick advective disc with $M_6 \sim 1$ and $B_S \sim 1$
 will always be accreting at a rate that implies  that the Eddington limit is exceeded.

We  further make the assumption that the accretion rate may be taken to be that provided by the stream so that  $\dot M \sim \dot M_S.$\footnote{Note that this assumption may not be valid
when $\alpha $ is small, however, we have checked that it holds for the considered
values of $\alpha, $ }
 We recall that $\dot M_{E}\approx 1.7\cdot 10^{23}M_6g/s$
and use equation (\ref{eq4}) to find that in order
the inequality (\ref{en2}) to be broken we should require that
\begin{equation}
\tau < \tau_{crit} \sim 20 \left ({B_SB_p^3\over \delta }\right)^{3/5}M_6^{-1/5}.
\label{en2n}
\end{equation}
Note that the parameters used in our simulation, $\delta \sim 0.5$, $B_S=0.5B_p\approx 1.55$ and $M_6=1$
equation (\ref{en2n}) gives $\tau_{crit}\approx 60$, which is somewhat smaller than the numerically obtained value
$\tau_{crit}\approx 80-100$. This, however, is well within an expected uncertainty based on the approximations
leading to (\ref{en2n}). Also, disc's semi-thickness $\delta $ is smaller than $0.5$ in the end of high
state, which could lead to larger $\tau_{crit}$ in a more accurate model.
It is instructive to rewrite (\ref{en2n}) in physical units.  With the  help of (\ref{eq3}) we obtain
\begin{equation}
t_{crit}\equiv P_{min}\tau_{crit}\approx 2(B_S/\delta)^{3/5}B_p^{-6/5}M_6^{7/10}yr.
\label{en3}
\end{equation}
Thus, the expected time to  the beginning of the transition to the low state, and, accordingly, the most
prominent deformation of the disc shape due to the stream influence is  of the order of a few months to years.

\subsection{Conditions for misalignment}
Now let us consider conditions under which the inclination angle $\beta $ can attain values order of $\beta_*$
at $\tau_{crit}$. As we have seen in Section \ref{gov} and the Appendix A, the important quantities determining the degree of misalignment
are the ratio of the alignment radius to $R_S,$   $\eta$ or $\eta_{rel}$ together with  $\sigma$ and $\sigma_{rel}$ respectively.
The latter quantities measures the ratio of the warp diffusion time or propagation time  to the accretion time.
Accordingly a large value for them, coupled with  small or modest values for the ratio of the alignment radius to $R_S,$ favours misalignment (see Appendiix A).

We  begin by evaluating typical values of the parameter $\sigma$ given by equation (\ref{eq14})
and  accordingly, of the parameter $\sigma_{rel}$ in the two states, see equation (\ref{eq26}). In order to do this we assume that
during the high state an approximate equilibrium between mass flow in the stream and  disc is maintained with
the mass flow in the disc being  close to the typical value $\dot M=3\pi \nu \Sigma $.
Equating this expression
to $\dot M_S$ and using equations (\ref{eq4}) and (\ref{eq5}) we find
\begin{equation}
\tilde  \Sigma_{high} \sim {1\over 3}\tau^{-5/3}/\left(
\alpha \delta_{high}^2 P_{min}\Omega_{S}\right),\label{sighigh}
\end{equation}
 where $\delta_{high}\sim 1$ is a typical disc  semi-thickness during
the high state.

When $\tau \sim \tau_{min},$ $\delta$ drops to its low value $\delta_{low}\sim 10^{-3}$. At this
stage the equilibrium between mass flows in the disc  stream is no longer possible, since a typical
viscous relaxation time in the disc becomes very long, and the disc is not able to effectively transfer the
mass supplied by the stream from the stream impact region. Therefore, there is an accumulation of mass near
$R\sim R_S$. However,  at the  beginning of this stage the mass
accumulated in this region will be rather small and we can roughly estimate the surface density at this time 
to be equal to its value during the preceding high state.   Substituting this value which  may be obtained from
(\ref{sighigh})  into   (\ref{eq14})
and remembering that $\delta =\delta_{low}$ we get an estimate for the  value of $\sigma$ in the high state as
\begin{equation}
\sigma_{max}\sim 6\left({\delta_{high}\over \delta_{low}}\right)^{2}\alpha^{2}\approx 13.5q\alpha_{-2}^2,
\label{en4}
\end{equation}
where $\alpha_{-2}=\alpha/10^{-2}$, and $q=({\delta_{high}/(150\delta_{low}}))^{2}$. Note that in our estimate
\noindent of the typical $q$ we assume that the disc semi-thickness just before  downward transition is order of $0.15$
as suggested in our numerical simulations, see Fig. \ref{Fig8}.  Following the same procedure to estimate  $\sigma_{rel}$ in the high state,
  we obtain from (\ref{eq26})
\begin{equation}
\sigma_{rel, max}\approx 100B_SM_6^{2/3}q\alpha_{-2}.
\label{en5}
\end{equation}
We remark that for $\alpha_{-2} > \sim 10$ as in our simulations, both  $\sigma_{max}$ and $\sigma_{rel,max}$
are on the order of $1000$ which, as indicated above, is favourable for misalignment.


In order to estimate values of the inclination angle we use the results of Appendix A together with expressions (\ref{eq13})
and (\ref{eqn16}) with $\delta=\delta_{low}=10^{-3},$ and also equations (\ref{en4}) and (\ref{en5}). Since
the results of Appendix A indicate little difference between prograde and retrograde black hole
rotation,  we do not distinguish between   these here.

 Expressed in terms of variables used in this
Section,  the  quantities  $\eta $ and $\eta_{rel},$ which measure the ratio of the alignment radius to $R_S$  and so  also  characterise the disc tilt,  take the form $\eta=40B_S(\alpha_{-2} a M_6)^{2/3}$
and $\eta_{rel}=20B_S a^{2/5}M_6^{2/3}$, where we set $a=|a|$ from now on.  Note that these quantities are always small
for small $a.$ From the discussion given in Appendix A, this implies that there will then always be misalignment.

Assuming that $\alpha~ >~\alpha_{crit}~=~4\cdot~10^{-3}~a^{-2/5},$  we can use equation (\ref{eqn22}), to obtain
\begin{equation}
\beta \sim \left({\sigma _{max}\eta^{-3/4}\over 2\sqrt{3}}\right)\beta_*\approx 0.2B_S^{-3/4}q\alpha_{-2}^{3/2}(aM_6)^{-1/2}\beta_*,
\label{en6}
\end{equation}
when the expression in brackets in smaller than $1$,  with  $\beta \sim \beta_{*}$ otherwise. The latter condition
can be expressed in the form
\begin{equation}
\alpha_{-2} > \sim 3B_{S}^{1/2}q^{-2/3}(aM_{6})^{1/3}.
\label{en7}
\end{equation}

When $\alpha_{-2}$ is smaller than $0.4,$  it is the relativistic correction $k$ in equation (\ref{eq12}), which determines
the shape of the disc and we should use the results of Appendix A devoted to the
case $\alpha < \alpha_{crit}$. These are analogous to  those of the previous case, see equation (\ref{eqn26}), but now $\sigma_{max} $
and $\eta $ must be substituted by $\sigma_{rel, max}$ and $\eta_{rel}.$  Using equations (\ref{eqn26})
and(\ref{eq30}) we obtain
\begin{equation}
\beta \approx 0.2\sigma _{rel, max}\eta_{rel}^{-5/4} \beta_* \approx 0.5 B_{s}^{-1/4}q\alpha_{-2}a^{1/2}M_6^{-1/6}\beta_*,
\label{en8}
\end{equation}

\noindent when $\sigma_{rel, max} < 5\eta_{rel}^{5/4}$, with $\beta \sim \beta_{*}$ otherwise. When  $\sigma_{rel, max} > 5\eta_{rel}^{5/4}$
we have
\begin{equation}
\alpha_{-2} > 2 B_S^{1/4}q^{-1}a^{1/2}M_6^{1/6}.
\label{en9}
\end{equation}

Equations (\ref{en7}) and (\ref{en9}) show  that in order to have $\beta \sim \beta_*$ at $\tau \sim \tau_{crit}$
either the viscosity parameter $\alpha $ should be larger than, say, $10^{-2}$, or the black hole rotation parameter
should be small enough. Let us stress that the fact that the inclination angle being close to that of the stream does not
mean that the disc is flat. It has a twisted form as long as,  either $R_{BP},$ in case of $\alpha > \alpha_{crit},$ or
$R_{rel} $ in the opposite case are larger than the size of marginally stable orbit. These conditions typically hold
for accretion discs in the low state unless the rotational parameter is very small.

\section{Discussion}
\label{Disc}

In this Paper we have considered  the influence of a stream of gas  acting as a source of mass for an accretion
disc around a rotating black hole on the  geometrical shape of the disc.
Both the accretion disc and the stream are assumed to have  originated
from the tidal disruption of a star by the black hole. The action of the gravetomagnetic force
tends to drag the disc towards the black hole equatorial plane, while the action of the stream
is  to  cause disc material  in the vicinity of the stream impact radius, $R_S$, to align its angular momentum vector with
that of the  orbit the star moved in prior to disruption.

Since, in general, this orbital plane is inclined with respect to the equatorial plane of the black hole by some angle order of unity,
 the combined action of the stream and the  gravitomagnetic force due to the black hole could result in the formation of a twisted disc.
 The presence of a tilted and twisted disc  could  have important observational consequences.
 In particular, it could affect the  spectrum and  produce variability \citep[e.g.][]{Dex2013},
  produce distinctive features in emission-line profiles \citep[e.g.][]{Bac1999},   provide a mechanism for the  excitation of quasi-periodic oscillations  through a parametric instability \citep[e.g.][]{Fer2008, Fer2009}.
  In turn, measurements of  disc  twist and tilt  could help to estimate both the black hole  parameters (its mass and spin)
as well as quantities governing the evolution of the  accretion disc.

We have employed  both analytic and numerical methods in our study. First, we extended the linear theory
of a stationary twisted disc to incorporate  an additional source of angular momentum provided
by the stream in Sections \ref{Basic}  - \ref{Torq} . We used this to identify the important parameters governing the shape of the disc in Section \ref{important}.
These were related to the ratio of the warp diffusion or propagation time to the mass accretion time
and the ratio of the alignment radius to the stream impact radius.  By considering both numerical solutions of the governing equation in Section \ref{Numsol}
as well as an analytic approach that yielded the asymptotic dependence of the disc tilt on these quantities given in an appendix,
 it was demonstrated a large value of the first  parameter and a small value of the second favoured the  misalignment of  a quasi-stationary disc.

We then  used SPH simulations to test these results  for a locally isothermal disc with aspect ratio $\sim 0.1$ in Section \ref{SPH}.
Near disc alignment for black hole rotation parameter, $a=1,$ and significant misalignment for $a=0.1$ was obtained for both approaches.
We also found reasonable  agreement for the  relaxed values of the
disc inclination angle , $\beta$,  at the stream impact radius, $R_{S},$ with a typical difference between analytic and SPH results being  $\sim 30\%.$

In order to generate more realistic models,  we  went on to use a one-dimensional  grid based  numerical scheme
to calculate the evolution of  background model discs  taking into account both gas and radiation pressure
in Section \ref{Grid}. Quantities such as  the disc surface density, $\Sigma$  and $\delta $ entering our equation for the disc tilt
were evolved forwards in time under the assumption that the  effective viscosity in the disc is described by the standard $\alpha$ model.
 The influence of the stream is taken into account through  a mass source  term localised  in radius to the vicinity
of $R_S.$  The  total mass flux was set  equal  to that of the stream.
In this way we obtained a sequence of background models for $\alpha =0.3$ and $\alpha =0.1$ without tilt,  spanning times between
the initial TDE and a  time at which most of the disc is in a cool state.
Thus the models were  evolved through a slim  disc  advection dominated stage and
the beginning of the transition to the radiative stage.
 It is important to stress that the  $\alpha $ model predicts that this transition is accompanied by thermal instability \citep[][]{Sha1976},
 which leads to limit cycle like behaviour  between 'high' and 'low' states of the disc with $\delta \sim 1$ and $0.001$, respectively.

Using these background models we calculated a sequence of quasi-stationary twisted disc configurations
and so found the dependence of $\beta$ on time in Section \ref{Gridmodels}.
We found that when the black hole rotation is close to the maximal one,
$\beta \sim 0.1$ of the stream inclination, $\beta_*$, while it grows to $\sim 0.4\beta_*$
when the disc experiences  transitions to low states. For smaller black hole rotations
these inclinations are larger.

With help of the asymptotic  analytic theory of solutions to our twisted disc equations outlined
in the appendix, we estimate
$\beta $ at the low state for accretion discs with  smaller values of $\alpha$  than those
adopted above. Thus $\beta $ is found to be $\sim 0.1\beta_*$ when $\alpha \sim 0.01$
and the  black hole rotation is close to  maximal. The disc inclination gets smaller for
smaller values of $\alpha $ and increases with decrease of the black hole rotation parameter.

That the disc changes its inclination during the transition to the low state could have  significant
observational consequences. If the orientation of the disc with respect to the line of sight is
such that parts of the disc with radii order of $R_S$ obscure the central source when it is
in the low state, there could be intermittent dips in its luminosity.
Note that such dips
have indeed been observed in a candidate TDE \citep[][]{Liu2014}, although  they  have been given a different interpretation
as being formed due to deflection of the stream by the gravitational field of another black
hole orbiting the one that produced the TDE \citep[][]{Liu2009, Liu2014}. If the  disc's orientation is such that both inner and
outer parts of it can be observed during high and low states,  the transitions could have
impact on the radiation spectrum.
In particular, when parts of the disc at $R\sim R_S$ are
inclined with respect to inner ones, they can easily intercept radiation coming from the central
source. Disc gas being heated up by a strong flux of X-rays could also form additional features
in the radiation spectrum.

Note that a strong increase of $\beta $ during a transition to the low state is expected even
for disc  models where the thermal instability does not operate. Unlike the models considered
in this Paper,  in the latter case,  this transition happens only once, and the subsequent alternation
of $\beta $ between relatively large and small values is not expected.

It is important to point out 
that the results of this Paper should be viewed as  first estimates. They have been
obtained under a number of significant simplifications. Perhaps, the most crucial one would be
our assumption that the disc is quasi-stationary. This holds well during the advection dominated
stage when timescales associated with relaxation to a quasi-stationary state are shorter than the
characteristic evolution timescale of the system. However, the relaxation time scale greatly increases
during the low state.
A simple estimate indicates that it is  significantly longer than the  evolution time when transitions between high and low states occur.
Clearly, in such a situation  time-dependent
calculations of the evolution of the disc tilt and twist are important.
The quasi-steady disc structures we have calculated,  at any time, can be viewed as
targets that  the system may not realise. Accordingly, the range of oscillations
in $\beta$ may be reduced in comparison to expectations from estimates
 made from quasi-stationary models. On the other hand misalignment is expected to still remain significant.

Note too that during the advection dominated stage there are additional terms in the twist equation, which have
been omitted, for simplicity, in this Paper. One should also take into account the possibility of strong outflows
during this stage from the disc. 
The analysis of a more complicated twist equation,
which has  explicit time dependence and at least partially accounts for effects determined
by advection will be the subject of a separate study.

\section*{Acknowledgements}
M. Xiang-Gruess acknowledges support through Leopoldina fellowship programme (fellowship number LPDS 2009-50).
Simulations were performed using the Darwin Supercomputer of the University of Cambridge High Performance Computing Service, provided by Dell Inc. using Strategic Research Infrastructure Funding from the Higher Education Funding Council for England and funding from the Science and Technology Facilities Council. Xiang-Gruess acknowledges the computing time granted (NIC project number 8163) on the supercomputer JUROPA at J\"ulich Supercomputing Centre (JSC) .
P. B. Ivanov was supported in part by RFBR grants 15-02-08476 and 16-02-01043 and also by Grant of the President of the Russian Federation for Support of the Leading Scientific Schools NSh-6595.2016.2.


\begin{appendix}

\section{Approximate analytic solutions to the  governing equation with estimates of $\beta_S$
for large and small  alignment radii }

When the source term in (\ref{eq12}) is set to zero  and $\delta $ and $\xi$ are constants
it can be solved exactly in two limiting cases $k \rightarrow 0$
and $k \rightarrow \infty $, see e.g. II. These cases correspond to the dominance of either viscous terms or post-Newtonian
relativistic corrections in equations of motion describing the twisted disc, respectively. Its dynamics and the shape
of stationary configurations differ qualitatively in these limits, in particular, as we have mentioned above when the relativistic corrections dominate and the black hole rotates in the same sense as the disc the disc's alignment with
the equatorial plane at small radii is absent.

When the source term is present we can also find solutions in the same limits
assuming that $\Delta \rightarrow 0$ in (\ref{eq11}) and, accordingly
 $\delta_{\Delta}$ reduces to the Dirac delta function.
Then the presence of the source term in (\ref{eq12}) results in
 a jump condition for the derivative of ${\cx W}$ at $r=1.$ This
is easily obtained by integrating   (\ref{eq12}) over an infinitesimal radial  domain centred on  $r=1.$  Assuming that ${\cx W}$
 is continuous we obtain
\begin{equation}
{d{\cx W}_{+}/dr} -{d {\cx W}_{-}/dr} =-\sigma (1-ik)({\cx W}_*-{\cx W}),
\label{eq15}
\end{equation}
where $d{\cx W}_{\pm}/dr$ are respectively  derivatives of solutions to (\ref{eq20})
evaluated for  $r > 1$ and $r < 1$ as $r\rightarrow 1.$
Thus, in our analytic approach a solution to (\ref{eq12}) consists of
homogeneous solutions in the inner domain ($r < 1$) and  the outer domain ($r > 1$) which satisfy  appropriate boundary conditions and
are linked through the jump condition (\ref{eq15}).

\subsection{The case $\alpha > \alpha_{crit}$}

We begin by  considering  the case where  viscous effects dominate and set $k=0$ in (\ref{eq12}). In this case
the source free solutions can be expressed  in the form
\begin{equation}
{\cx W}=z_1^{1/3}(C_1J_{-1/3}(z_1)+C_2J_{1/3}(z_1)), \quad z_1={4\over 3}c_{\pm}e^{i\pi/4}\left({\eta \over r}\right)^{3/4},
\label{eq17}
\end{equation}
where $C_{1,2}$ are arbitrary constants, $J_{\nu}(z)$  is the  Bessel functions of order $\nu,$
the  $(+)$ and $(-)$  signs  respectively correspond
to prograde and retrograde black hole rotation,  $c_{+}=1$ and  $c_{-}=i$.

The solution (\ref{eq17}) should satisfy two boundary conditions, $|{\cx W}| \rightarrow 0$ when $r \rightarrow 0$,
and ${\cx W}^{'}\rightarrow 0$ when $r\rightarrow \infty $, and  hereafter a prime denotes  the  radial derivative .
 The inner  solution, valid for $r < 1,$  that satisfies the inner boundary condition can be written in the form
\begin{equation}
{\cx W}_{in}=C_{in}\phi_{in},\hspace{2mm}  {\rm with} \quad \phi_{in}=z_1^{1/3}(J_{-1/3}(z_1)-e^{-i\pi/3}J_{1/3}(z_1)),
\label{eq18}
\end{equation}
while the outer  solution valid for  $r > 1$ that satisfies the outer boundary condition  may be written as
\begin{equation}
{\cx W}_{out}=C_{out}\phi_{out}, \hspace{2mm}  {\rm with} \quad \phi_{out}=z_1^{1/3}J_{-1/3}(z_1).
\label{eq19}
\end{equation}
The jump condition (\ref{eq15}) can be used to find $C_{in}$, $C_{out}$ and accordingly, the value of ${\cx W}_{S}~=~{\cx W}~(~r~=~1~)$.
The calculation is straightforward with the result that
\begin{equation}
{\cx W}_{S}={\sigma f {\cx W}_*\over \sigma f +1},\hspace {2mm}{\rm where}  \quad f=-\phi_{in}\phi_{out}/W_R, \hspace {2mm}{\rm with\hspace{1mm}  
the \hspace{1mm} Wronskian} \quad W_R=\phi_{in}\phi_{out}^{'}-\phi_{out}\phi_{in}^{'}.
\label{eq20}
\end{equation}
with all  quantities being evaluated at $r=1.$

With no loss of generality we can assume that ${\cx W}_*$ is real and
equal to $\beta_*$. Then   $\beta_S$   can be found from (\ref{eq20}) in the form
\begin{equation}
\beta_{S}={\sigma |f| \beta_*\over \sqrt{D}}, \hspace{2mm} {\rm where} \quad D=|f|^2\sigma^2+
(f+f^*)\sigma+1.
\label{eqn20}
\end{equation}
From the known properties of Bessel and Gamma functions we  obtain
\begin{equation}
W_R= -{2^{-2/3}3^{5/6}c_{\pm}^{2/3}\over \pi}e^{-i\pi/6}\eta^{1/2}r^{-3/2}.
\label{eq21}
\end{equation}
\subsubsection{Expressions for $\beta_S $ in the limits of large and small $\eta$}
Using  (\ref{eq21}) together with asymptotic expressions for the Bessel functions, we can express    $f$
in (\ref{eq20}) in terms of elementary functions in the limits $\eta \gg 1$ and $\eta
\ll 1$.
 In the former limit we can use the asymptotic forms of  the Bessel functions for large absolute values of their
arguments to find
\begin{equation}
f={e^{i\pi/12}(1-e^{-i\pi/3})\over 2\sqrt 3 c_{\pm}}\eta^{-3/4}.
\label{eq22}
\end{equation}
When the black hole rotation is prograde we can set $c_{\pm}=1$ and obtain from
(\ref{eq22})
\begin{equation}
\beta_S={\sigma \eta^{-3/4}\over 2\sqrt {D}}\beta_*, \hspace {2mm} {\rm where} \quad
D={\eta^{-3/2}\over 4}\sigma^2+{\sqrt{(\sqrt 3+2)}\eta^{-3/4}\over 2}\sigma
+1.
\label{eqn22}
\end{equation}
When $a < 0$ the expression for $\beta_{S}$ is the same as (\ref{eqn22})
with the  modification that  as in this  case $f$ is purely imaginary,
 the term proportional to $\sigma $ in the expression for $D$ is
equal to zero as can easily be seen from (\ref{eqn20}). Since the contribution of this term is quite small we
use equation (\ref{eqn22}) for both prograde and retrograde rotation.

When $\eta \ll  1$ we use the asymptotic representations of Bessel function at small absolute values of their
arguments to obtain
\begin{equation}
f=\left({2^{4/3}\pi e^{{\rm i}\pi/6}\over 3^{5/6}\Gamma^{2}(2/3)c_{\pm}^{2/3}}\right)\eta^{-1/2}
\approx 1.17\left({e^{{\rm i}\pi/6}\over c_{\pm}^{2/3}}\right)\eta^{-1/2}.
\label{eq23}
\end{equation}
Since  $|f|$ is the same for both prograde and retrograde
cases, for  both  we approximately have
\begin{equation}
\beta_S={1.17\sigma \eta^{-1/2}\beta_*\over \sqrt D}, \quad D=1.3689\eta^{-1}\sigma^2
+ 1.17\sigma \sqrt{3\eta^{-1}}+1.
\label{eqn23}
\end{equation}
Taking the limit $\eta \rightarrow 0$ we obtain $\beta_S \rightarrow  \beta_*.$
In this case, although warp propagation is efficient,  the alignment radius  is arbitrarily small
so that disc is unable to align with the black hole equator.
As $\eta$ increases the degree of misalignment decreases, being governed by
the magnitude of the  quantity $\sqrt{\eta}/\sigma.$ When $R_{BP}=R_S,$  this is proportional to the ratio
of the local accretion time  scale to the local warp diffusion time scale.
A small value of this ratio favours misalignmemt as expected.

On the other hand in the opposite limit $\eta \rightarrow \infty,$ corresponding to the alignment radius
moving to large radii, $\beta_S \rightarrow 0$ as  expected. As before a large value of $\sigma$
tends to favour misalignment.

Note that the case $\alpha > \alpha_{crit}$ and $\eta < 1$ can be realised
only when the black hole mass is sufficiently small if we require that $\alpha < 1.$
Indeed, from these conditions we get

\noindent   $\alpha_{crit}~<~\alpha~<~({23/ 2})^{3/2}\delta^2 (a \beta_S M_6)^{-1}$,
which leads to  $a < ({23/ 2})^{5/2}\delta^2 (B_S M_6)^{-5/3}$.
Using  again the fact that here we are considering   $\alpha > \alpha_{crit}=a^{-2/5}\delta^{4/5},$
we obtain from the above condition on,  $a,$ that $\alpha~<~{2/ 23}(B_S M_6)^{2/3}$.
 Since, on the other hand $\alpha $ is required to be smaller than
one we must have $M_6 < \sim{39/B_S}$.

\subsection {The case $\alpha < \alpha_{crit}$}

In this case we  formally assume that the constant $k$ in (\ref{eq12}) becomes very large. In this limit the homogeneous
solutions to (\ref{eq12}) can be written as
\begin{equation}
{\cx W}=z_2^{3/5}(C_1 J_{-3/5}(z_2)+ C_2 J_{3/5}(z_2)), \hspace{2mm} {\rm where} \quad z_2={2\sqrt{3}\over  5c_{\pm}}\left({\eta_{rel}\over r}\right )^{5/4},
\label{eq24}
\end{equation}
and  we recall  that $\eta_{rel}$ is given by equation (\ref{eqn16}). As for the previous case we introduce inner and outer solutions according to the specification
\begin{equation}
{\cx W}(r <1)=C_{in}\phi_{in}, \hspace{2mm} {\rm and} \quad {\cx W}(r > 1)= C_{out}\phi_{out},                   \label{eq25}
\end{equation}
where $\phi_{in}$ and $\phi_{out}$ are proportional to   combinations of Bessel functions discussed below.
The expression (\ref{eq20}) is modified by the substitution $\sigma \rightarrow -ik\sigma $:
\begin{equation}
{\cx W}_{S}={{\rm i} \sigma_{rel} f {\cx W}_*\over {\rm i}\sigma_{rel} f +1}, \quad \sigma_{rel}=k\sigma = {12\lambda GM\over \delta^2 c^2 R_S}
\left({\dot M_S\over 2\pi \Sigma R_S^2}\sqrt{{R_S^3\over GM}}\right).
\label{eq26}
\end{equation}
As we will see below that now $f$ is real and positive in both limits of large and small $\eta_{rel}$. Using this fact we easily obtain from (\ref{eq26}) that
\begin{equation}
{\beta}_{S}={f\sigma_{rel}\over \sqrt {f^2\sigma_{rel}^2 + 1}}\beta_*.
\label{eqn26}
\end{equation}
Note that $\sigma_{rel}$ plays the role of $\sigma$ in this case and may be regarded
as being equal to $\sigma$ with $\alpha$ set equal to $3\lambda GM/(c^2R_S^2).$

As we have discussed above when $\alpha < \alpha_{crit}$ there is
a qualitative difference in behaviour of solutions  of the governing corresponding to
different signs of $a$ (see section \ref{gov}). Therefore,
it is convenient to treat them separately.

\subsubsection{The case of prograde rotation $a > 0$}

We recall  that when $a > 0,$  $z_2$ is real and the inner oscillating solution is characterised by a phase, $\Psi$, which is determined
by conditions close to the last stable orbit, where one must consider a fully relativistic theory of twisted discs, see
ZI for such an approach. Here we shall fix this phase, $\Psi,$
to be equal to $-2\pi/3$, which is obtained in a certain limit discussed in ZI. In order to comply with the  notation of this paper we write the inner solution in the form
\begin{equation}
\phi_{in}=z_2^{3/5}(\cos (\Psi - \pi/20)J_{-3/5}(z_2)+\sin (\Psi +\pi/20)J_{3/5}(z_2)),                                                           \label{eq27}
\end{equation}
while the outer solution is given by  $\phi_{out}=z_2^{3/5}J_{-3/5}(z_2)$. The Wronskian $W_R$ of this pair can be written as
\begin{equation}
W_R=({54/ 5})^{1/5}\pi^{-1}{\sin ({2\pi/ 5})\sin (\Psi + \pi/20)}\eta_{rel}^{3/2}r^{-5/2}.                                                     \label{eq28}
\end{equation}
When $\eta_{rel} \gg 1$ we approximately have
\begin{equation}
f={2\over \sqrt  3}{\cos (z_S-\Psi)\cos (z_S+\pi/20)\over \sin (\Psi + \pi/20)}\eta_{rel}^{-5/4},                                                         \label{eq29}
\end{equation}
where $z_S=z_2(r=1)={2\sqrt{3}/5}\eta_{rel}^{5/4}$.

\noindent
In order to obtain  an order of magnitude estimate of $\beta_S$ we further
simplify (\ref{eq29}) by averaging over $z_S$ to obtain
\begin{equation}
f={1\over \sqrt  3}\tan^{-1} (\Psi+\pi/20)\eta_{rel}^{-5/4}\approx 0.22\eta_{rel}^{-5/4},                                                   \label{eq30}
\end{equation}
where we assume $\Psi=-2\pi/3$ to obtain the last equality.

In the opposite limit $\eta_{rel} \ll 1$ we get
\begin{equation}
f=({5/ 27})^{1/5}\left({2\pi \over \Gamma^2 (2/5)}\right)\left({\cos (\Psi - \pi /20)\over \cos (\pi/10)\sin (\Psi +\pi/20)}\right)
\eta_{rel}^{-3/2}\approx 0.645  \eta_{rel}^{-3/2},                                               \label{eq31}
\end{equation}
where  we have again set $\Psi=-2\pi/3$ in the last equality.

As expected when the alignment radius approaches zero with $\eta_{rel}\rightarrow 0,$ we obtain $\beta_S \rightarrow \beta_*$
corresponding to complete misalignment. Similarly in the opposite limit with $\eta_{rel}\rightarrow \infty,$
we obtain $\beta_S\rightarrow 0,$ corresponding to complete alignment. The magnitude of the  quantity $\sigma_{rel}$
then determines the degree of misalignment for a given $\eta_{rel}$  through equation (\ref{eqn26}), with a large value favouring misalignment.

\subsubsection{The case of retrograde rotation $a < 0$}

When $a < 0,$ $z_2$ is purely imaginary and the inner solution is fixed by the requirement that the disc inclination
should tend to zero at small $r$. We then have
\begin{equation}
\phi_{in}=z_2^{3/5}(J_{-3/5}(z_2)-e^{-i3\pi/5}J_{3/5}(z_2)), \hspace{2mm} {\rm and} \quad  \phi_{out}=z_2^{3/5}J_{-3/5}(z_2),                                                  \label{eq33}
\end{equation}
which lead to
\begin{equation}
W_R=-({54/ 5})^{1/5}\pi^{-1}{\cos ({\pi/ 10})}\eta_{rel}^{3/2}r^{-5/2}.                                                      \label{eq34}
\end{equation}
When $\eta_{rel} \gg 1$ we find from equation (\ref{eq20})  that
\begin{equation}
f={1\over \sqrt 3}\eta_{rel}^{-5/4}\approx 0.58\eta_{rel}^{-5/4},                                                              \label{eq35}
\end{equation}
and when
$\eta_{rel} \ll 1$ we obtain
\begin{equation}
f=2\pi({5/  27})^{1/5} \left( \cos (\pi /10)\Gamma^2(2/5)\right)^{-1}\eta_{rel}^{-3/2}\approx 0.96\eta_{rel}^{-3/2}.                                               \label{eq36}
\end{equation}
From  equation (\ref{eqn26}) and  the expressions (\ref{eq35}) and (\ref{eq36}) it follows that the estimate of $\beta_S$ for  the retrograde case will be to
order of magnitude the same as for the prograde case.

\end{appendix}

\label{lastpage}


\begin{thebibliography}{99}

\bibitem[\protect\citeauthoryear{Abramowicz et al.}{1988}]{Abr1988} Abramowicz M. A., Czerny B., Lasota J. P., Szuszkiewicz E., 1988, ApJ, 332, 646
\bibitem[\protect\citeauthoryear{Bachev}{1999}]{Bac1999} Bachev R., 1999, A \&A, 348, 71
\bibitem[\protect\citeauthoryear{Bardeen \& Petterson}{1975}]{Bar1975} Bardeen J. M. \& Petterson J. A., 1975, ApJ, 195, L65
\bibitem[\protect\citeauthoryear{Bogdanovic et al.}{2004}]{Bog2004} Bogdanovic T., Eracleous M., Mahadevan S., Sigurdsson S., Laguna P., 2004, ApJ, 610, 707
\bibitem[\protect\citeauthoryear{Bonnerot et al.}{2016}]{Bon2016} Bonnerot C., Rossi E. M., Lodato G., Price D. J., 2016, MNRAS, 455, 2253
\bibitem[\protect\citeauthoryear{Burrows et al.}{2011}]{Bur2011} Burrows D. N., Kennea J. A., Ghisellini G., Mangano V., Zhang B., Page K. L., Eracleous M., Romano P., Sakamoto T., Falcone A. D., Osborne J. P., Campana S., Beardmore A. P., Breeveld A. A., Chester M. M., Corbet R., Covino S., Cummings J. R., D'Avanzo P., D'Elia V., Esposito P., Evans P. A., Fugazza D., Gelbord J. M., Hiroi K., Holland S. T., Huang K. Y., Im M., Israel G., Jeon Y., Jeon Y.-B., Jun H. D., Kawai N., Kim J. H., Krimm H. A., Marshall F. E., P. M{\'e}sz{\'a}ros, Negoro H., Omodei N., Park W.-K., Perkins J. S., Sugizaki M., Sung H.-I., Tagliaferri G., Troja E., Ueda Y., Urata Y., Usui R., Antonelli L. A., Barthelmy S. D., Cusumano G., Giommi P., Melandri A., Perri M., Racusin J. L., Sbarufatti B., Siegel M. H., Gehrels N., 2011, Nature, 476, 421
\bibitem[\protect\citeauthoryear{Cannizzo et al.}{1990}]{Can1990} Cannizzo J. K., Lee H. M., Goodman J., 1990, ApJ, 351, 38
\bibitem[\protect\citeauthoryear{Caproni et al.}{2007}]{Cap2007} Caproni A., Abraham Z., Livio M., Mosquera Cuesta H. J., 2007,  MNRAS, 379, 135
\bibitem[\protect\citeauthoryear{Carter \& Luminet}{1983}]{Car1983} Carter B. \& Luminet J.-P., 1983, A$\&$A, 121, 97
\bibitem[\protect\citeauthoryear{Carter \& Luminet}{1985}]{Car1985} Carter B. \& Luminet J. P., 1985, MNRAS, 212, 23
\bibitem[\protect\citeauthoryear{Coughlin et al.}{2016}]{Cou2016} Coughlin E. R., Nixon C., Begelman M. C., Armitage P. J., Price D. J., 2016, MNRAS, 455, 3612
\bibitem[\protect\citeauthoryear{Demianski \& Ivanov}{1997}]{Dem1997} Demianski M. \& Ivanov P. B., 1997, A\&A, 324, 829
\bibitem[\protect\citeauthoryear{Dexter \& Fragile}{2013}]{Dex2013} Dexter J. \& Fragile P. C., 2013, ApJ, 730, 36
\bibitem[\protect\citeauthoryear{Dremova et al.}{2014}]{Dre2014} Dremova G. N., Dremov V. V., Tutukov A. V., 2014, Astronomy Reports,  58, 291
\bibitem[\protect\citeauthoryear{Esquej et al.}{2008}]{Esq2008} Esquej P., Saxton R. D., Komossa S., Read A. M., Freyberg M. J., Hasinger G.,  Garc{\'i}a-Hern{\'a}ndez D. A., Lu H., Rodriguez Zaur{\'i}n J., S{\'a}nchez-Portal M., Zhou H., 2008, A\&A, 489, 543
\bibitem[\protect\citeauthoryear{Evans \& Kochanek}{1989}]{Eva1989} Evans C. R. \& Kochanek C. S., 1989, ApJ, 346, L13
\bibitem[\protect\citeauthoryear{Ferreira \& Ogilvie}{2008}]{Fer2008} Ferreira B. T. \& Ogilvie G. I., 2008, MNRAS, 386, 2297
\bibitem[\protect\citeauthoryear{Ferreira \& Ogilvie}{2009}]{Fer2009} Ferreira, B. T. \& Ogilvie, G. I., 2009, MNRAS, 392, 428
\bibitem[\protect\citeauthoryear{Franchini et al.}{2015}]{Fra2015} Franchini A., Lodato G., Facchini S., 2015, MNRAS, 455, 1946
\bibitem[\protect\citeauthoryear{Frank \& Rees}{1976}]{Fra1976} Frank J. \& Rees M. J., 1976, MNRAS, 176, 633
\bibitem[\protect\citeauthoryear{Garavaglia}{1987}]{Gar1987} Garavaglia T., 1987, Am. J. Phys. 55, 164
\bibitem[\protect\citeauthoryear{Guillochon \& Ramirez-Ruiz}{2013}]{Gui2013} Guillochon J. \& Ramirez-Ruiz E., 2013, ApJ, 767, 25
\bibitem[\protect\citeauthoryear{Guillochon \& Ramirez-Ruiz}{2015}]{Gui2015} Guillochon J. \& Ramirez-Ruiz E., 2015, ApJ, 809, 166
\bibitem[\protect\citeauthoryear{Hayasaki et al.}{2013}]{Hay2013} Hayasaki K., Stone N., Loeb A., 2013, 434, 909
\bibitem[\protect\citeauthoryear{Hills}{1975}]{Hil1975} Hills J. G., 1975, Nature, 254, 295
\bibitem[\protect\citeauthoryear{Ivanov \& Illarionov}{1997}]{Iva1997} Ivanov P. B. \& Illarionov A. F.,  1997, MNRAS, 285, 394
\bibitem[\protect\citeauthoryear{Ivanov \& Novikov}{2001}]{Iva2001} Ivanov P. B. \& Novikov I. D., 2001, ApJ,  549, 467
\bibitem[\protect\citeauthoryear{Ivanov et al.}{2003}]{Iva2003} Ivanov P. B., Chernyakova M. A., Novikov I. D., 2003, MNRAS, 338, 147
\bibitem[\protect\citeauthoryear{Ivanov et al.}{2005}]{Iva2005} Ivanov P. B., Polnarev A. G., Saha P., 2005, MNRAS, 358, 1361
\bibitem[\protect\citeauthoryear{Ivanov \& Chernyakova}{2006}]{Iva2006} Ivanov P. B. \& Chernyakova M. A., 2006, A$\&$A, 448, 843
\bibitem[\protect\citeauthoryear{Ivanov et al.}{2015}]{Iva2015}Ivanov P. B., Papaloizou, J.C.B.,  Paardekooper, S.-J., Polnarev A. G., 2015, A\& A, 2015, 576, id.A29
\bibitem[\protect\citeauthoryear{Kelley et al.}{2014}]{Kel2014} Kelley L. Z., Tchekhovskoy A., Narayan R., 2014, MNRAS, 445, 3919
\bibitem[\protect\citeauthoryear{Khabibullin et al.}{2014}]{Kha2014} Khabibullin I., Sazonov, S., Sunyaev, R., 2014, MNRAS, 437, 327
\bibitem[\protect\citeauthoryear{Khokhlov et al.}{1993a}]{Kho1993a} Khokhlov A., Novikov I. D., Pethick C. J., 1993, ApJ, 418, 163, (a) 
\bibitem[\protect\citeauthoryear{Khokhlov et al.}{1993b}]{Kho1993b} Khokhlov A., Novikov I. D., Pethick C. J., 1993, ApJ, 418, 181, (b)
\bibitem[\protect\citeauthoryear{Kim et al.}{1999}]{Kim1999} Kim S. S., Park M.-G., Lee H. M., 1999, ApJ, 519, 647
\bibitem[\protect\citeauthoryear{Kochanek}{1994}]{Koc1994} Kochanek C. S., 1994, ApJ, 422, 508
\bibitem[\protect\citeauthoryear{Komossa et al.}{2008}]{Kom2008} Komossa S., Zhou H., Wang T., Ajello M., Ge J., Greiner J., Lu H., Salvato M., Saxton R., Shan H., Xu D., Yuan W., 2008, ApJ Letters, 678, L13
\bibitem[\protect\citeauthoryear{Komossa et al.}{2009}]{Kom2009} Komossa S., Zhou H., Rau A., Dopita M., Gal-Yam A., Greiner J., Zuther J., Salvato M., Xu D., Lu H., Saxton R., Ajello M., 2009, Ap, 701, 105
\bibitem[\protect\citeauthoryear{Komossa}{2015}]{Kom2015} Komossa S., 2015, Journal of High Energy Astrophysics, 7, 148
\bibitem[\protect\citeauthoryear{Lacy et al.}{1982}]{Lac1982} Lacy J. H., Townes C. H., Hollenbach D. J.,  1982, ApJ, 262, 120L
\bibitem[\protect\citeauthoryear{Lightman \& Shapiro}{1977}]{Lig1977} Lightman A. P., Shapiro S. L., 1977, ApJ, 211, 244
\bibitem[\protect\citeauthoryear{Liu et al.}{2014}]{Liu2014} Liu F. K., Li S., Komossa S., 2014, ApJ, 786, 103
\bibitem[\protect\citeauthoryear{Liu et al.}{2009}]{Liu2009} Liu F. K., Li S., Chen X., 2009,  ApJ, 706, L133
\bibitem[\protect\citeauthoryear{Lodato et al.}{2009}]{Lod2009} Lodato G., King A. R., Pringle J. E., 2009, MNRAS, 392, 332
\bibitem[\protect\citeauthoryear{Lubow et al.}{2002}]{Lub2002} Lubow S. H., Ogilvie G. I., Pringle J. E,  2002, MNRAS, 337, 706
\bibitem[\protect\citeauthoryear{Lynden-Bell \& Pringle}{1974}]{Lyn1974} Lynden-Bell D. \& Pringle J. E., 1974, MNRAS, 168, 603
\bibitem[\protect\citeauthoryear{MacLeod et al.}{2012}]{Mac2012} MacLeod M, Guillochon J., Ramirez-Ruiz E., 2012, ApJ, 757, 134
\bibitem[\protect\citeauthoryear{Magorrian \& Tremaine}{1999}]{Mag1999} Magorrian J. \& Tremaine S., 1999, 309, 447
\bibitem[\protect\citeauthoryear{Mainetti et al.}{2016}]{Mai2016} Mainetti D., Lupi A., Campana S., Colpi M., 2016, MNRAS, 457, 2516
\bibitem[\protect\citeauthoryear{Miller}{2015}]{Mil2015} Miller M. C., 2015, ApJ, 805, 83
\bibitem[\protect\citeauthoryear{Morales Teixeira et al.}{2014}]{Mor2014} Morales Teixeira D.,  Fragile P. C.,  Zhuravlev V. V., Ivanov P. B., 2014, ApJ, 796, 103
\bibitem[\protect\citeauthoryear{Nealon et al.}{2015}]{Nea2015} Nealon R., Price D. J., Nixon C. J., MNRAS, 2015, 448, 1526
\bibitem[\protect\citeauthoryear{Paczy{\'n}ski \& Wiita}{1980}]{Pac1980} Paczy{\'n}ski B. \& Wiita P. J., 1980, A\&A, 88, 23
\bibitem[\protect\citeauthoryear{Papaloizou \& Pringle}{1983}]{Pap1983} Papaloizou, J. C. B., Pringle, J. E., 1983, MNRAS, 202, 1181
\bibitem[\protect\citeauthoryear{Papaloizou \& Lin}{1995}]{Pap1995} Papaloizou, J. C. B., Lin, D. N. C., 1995, ApJ,  438, 841
\bibitem[\protect\citeauthoryear{Rees}{1988}]{Ree1988} Rees M. J., 1988, Nature, 333, 523
\bibitem[\protect\citeauthoryear{Shakura \& Sunyaev}{1973}]{Sha1973} Shakura N. I. \& Sunyaev R. A., 1973, A\&A, 24, 337
\bibitem[\protect\citeauthoryear{Shakura \& Sunyaev}{1976}]{Sha1976} Shakura N. I. \& Sunyaev R. A., 1976, MNRAS, 175, 613
\bibitem[\protect\citeauthoryear{Shen \& Matzner}{2014}]{She2014} Shen R.-F. \& Matzner C. D., 2014, ApJ, 784, 87	
\bibitem[\protect\citeauthoryear{Springel}{2005}]{Spr2005} Springel V. , 2005, MNRAS, 364, 1105
\bibitem[\protect\citeauthoryear{Stone \& Loeb}{2012}]{Sto2012} Stone N. \& Loeb A., 2012, Physical Review Letters, 108, 061302
\bibitem[\protect\citeauthoryear{Stone \& Metzger}{2016}]{Sto2016} Stone N. C. \& Metzger B. D., 2016, MNRAS, 455, 859
\bibitem[\protect\citeauthoryear{Syer \& Ulmer}{1999}]{Sye1999} Syer D. \& Ulmer A., MNRAS, 1999, 306, 35
\bibitem[\protect\citeauthoryear{Szuszkiewicz \& Miller}{1997}]{Szu1997} Szuszkiewicz E. \& Miller J. C., 1997, MNRAS, 287, 165
\bibitem[\protect\citeauthoryear{Szuszkiewicz \& Miller}{1998}]{Szu1998} Szuszkiewicz E. \& Miller J. C., 1998, MNRAS, 298, 888
\bibitem[\protect\citeauthoryear{Szuszkiewicz \& Miller}{2001}]{Szu2001} Szuszkiewicz E., Miller J. C., 2001, MNRAS, 328, 36
\bibitem[\protect\citeauthoryear{Thorne et al.}{1986}]{Tho1986} Thorne K. S., Price R. H.,  Macdonald D. M., 1986, Black Holes: The Membrane Paradigm (Yale University Press, New Haven, CT.)
\bibitem[\protect\citeauthoryear{Van Velzen et al.}{2011}]{Van2011} Van Velzen S., K{\"o}rding E., Falcke H., 2011, 417, L51
\bibitem[\protect\citeauthoryear{Wu et al.}{2010}]{Wu2010} Wu S.-M., Chen L., Yuan F., 2010, MNRAS, 402, 537
\bibitem[\protect\citeauthoryear{Zhang et al.}{2015}]{Zha2015} Zhang W., Yu W., Karas V., Dov{\~ c}iak, M., 2015, ApJ, 807, 89
\bibitem[\protect\citeauthoryear{Zhuravlev \& Ivanov}{2011}]{Zhu2011} Zhuravlev V. V. \& Ivanov P. B.,  2011, MNRAS, 415, 2122
\bibitem[\protect\citeauthoryear{Zhuravlev et al.}{2014}]{Zhu2014} Zhuravlev V. V., Ivanov P. B., Fragile P. C.,  Morales Teixeira, D., 2014, ApJ, 796, 104
\bibitem[\protect\citeauthoryear{Ziegler \& Yorke}{1997}]{Zie1997} Ziegler U. \& Yorke H. W., 1997, Computer Physics Communications, 101, 54

\end{thebibliography}
\end{document}